\documentclass[bibyear]{aa}  

\usepackage{graphicx}

\usepackage{txfonts}

\begin{document}

   \title{ Asymptotic g modes: Evidence for a rapid rotation of the solar core }
   
   \author{E.~Fossat\inst{1} \and
          P.~Boumier\inst{2} \and
           T.~Corbard\inst{1} \and J.~Provost\inst{1} \and D.~Salabert\inst{3} \and
           F.~X.~Schmider\inst{1} \and A.~H.~Gabriel\inst{2} \and  G.~Grec\inst{1} \and C.~Renaud\inst{1} \and J.~M.~Robillot\inst{4} \and T.~Roca-Cort\'es\inst{5,6} \and  S.~Turck-Chi\`eze\inst{7} \and R.~K.~Ulrich\inst{8} \and M.~Lazrek\inst{9} }

\institute{ Universit\'e  C\^ote d'Azur, Observatoire C\^ote d'Azur, CNRS, Laboratoire Lagrange,  CS 34229, Nice cedex 4, France \\ \email{eric.fossat@oca.eu}
\and
    Institut d'Astrophysique Spatiale, Universit\'e Paris-Sud and CNRS (UMR 8617), B\^atiment 121, F-91405 Orsay cedex, France
 \and
    Laboratoire AIM Paris-Saclay, CEA/DRF-CNRS-Univ. Paris Diderot - IRFU/SAp, Centre de Saclay, 91191 Gif-sur-Yvette, France
    \and
    LAB, 2 rue de l'Observatoire, BP89, 33271, Floirac Cedex, France
     \and
    Departamento de Astrof\'isica, Universidad de La Laguna, 38206 La Laguna,Tenerife Spain
    \and
    Instituto de Astrof\'isica de Canarias. 38205 La Laguna, Tenerife, Spain
    \and 
    DAP/IRFU/CEA, UMR AIM, University Paris-Saclay, CE Saclay, 91191 Gif sur Yvette, France
    \and
    Department of Physics and Astronomy Department, University of California at Los Angeles,
430 Portola Plaza, Box 951547, Los~Angeles, CA 90095-1547, USA
   \and
   LPHEA Laboratory, Oukaimeden Observatory, Cadi Ayyad University FSSM, BP 2390 Marrakech, Morocco 
                }
 
   \date{Received 18 January 2017 / accepted 24 May 2017} 
  \abstract
     {Over the past 40 years, helioseismology has been enormously successful in the study of the solar interior.  A shortcoming has been the lack of a convincing detection of the solar g modes, which  are oscillations driven by gravity and are hidden in the deepest part of the solar body -- its hydrogen-burning core. The detection of g modes is expected to dramatically improve our ability to model this core, the rotational characteristics of which have, until now, remained unknown. }
    {We present the identification of very low frequency g modes in the asymptotic regime and two important parameters that have long been waited for: the core rotation rate, and the asymptotic equidistant period spacing of these g modes. }
   {The GOLF instrument on board the SOHO space observatory has provided two decades of full-disk helioseismic data. The search for g modes in GOLF measurements has been extremely difficult
because of solar and instrumental noise. In the present study, the p modes of the GOLF signal are analyzed differently: we search for possible collective frequency modulations that are produced by periodic changes in the deep solar structure. Such modulations provide access to only very low frequency g modes, thus allowing statistical methods to take advantage of their asymptotic properties.  }
   { For oscillatory periods in the range between 9 and nearly 48 hours, almost 100 g modes of spherical harmonic degree 1 and more than 100 g modes of degree 2 are predicted. They are not observed individually, but when combined, they unambiguously provide their asymptotic period equidistance and rotational splittings, in excellent agreement with the requirements of the asymptotic approximations. When the period equidistance has been measured, all of the individual frequencies of each mode can be determined. Previously, p-mode helioseismology allowed the g-mode period equidistance parameter $P_{0}$ to be bracketed inside a narrow range, between approximately 34 and 35 minutes. Here, $P_{0}$ is measured to be 34~min 01~s, with a 1~s uncertainty. The previously unknown g-mode splittings have now been measured from a non-synodic reference with very high accuracy, and they imply a mean weighted rotation of  $1277  \pm 10$~nHz (9-day period) of their kernels, resulting in a rapid rotation frequency of  $1644\pm 23$~nHz (period of one week) of the solar core itself, which is a factor $3.8 \pm 0.1$ faster than the rotation of the radiative envelope.
}
   {The g modes are known to be the keys to a better understanding of the structure and dynamics of the solar core. Their detection with these precise parameters will certainly stimulate a new era of  research in this field.}

  \authorrunning {Fossat et al.}
 \titlerunning{Solar g modes}
 
 \maketitle


\section{Introduction}

A conference was organized in 2013 in Tucson, Arizona, to celebrate 50 years of seismology of the sun and stars. Solar seismology became known in the early 1980s as helioseismology, and was almost totally devoted in this conference to the analysis of solar p modes, sometimes referred to as solar music, as only one presentation among 87 was devoted to g modes. It  was a historical presentation  (Appourchaux \& Pall\'e 2013) that confirmed how difficult the quest for g modes has been in these 40 years.  

The observational  g-mode quest began 40 years ago, when Severny et al.~(1976) reported the detection of a 160-minute oscillation using a modified differential Babcock solar magnetograph and nine days of data. This period of 2 hours and 40 minutes could not be attributed to a p mode in any reasonable solar model. In 1976, the field was still in its infancy, with the phenomena being referred to as  solar oscillations and with no detections of individual p modes. Although the potential of p modes in probing the deep solar interior was still not clear, the potential of g modes was obvious, as these are mostly trapped inside the solar core. The 160-minute oscillation thus quickly became a hot topic, and at least  ten years passed before it finally became clear that it was an artifact.

Following the first identification of global p modes by Claverie et al.~(1979) and the first South Pole expedition, which allowed individual low spherical harmonic degree p modes to be revealed (Grec et al. 1980, 1983), helioseismology quickly grew on different fronts: on the ground with  world-wide networks that were developed by the Birmingham Solar Oscillation Network (BiSON) in the U.K., the International Research on the Interior of the Sun (IRIS) in France, and the Global Oscillation Network Group (GONG) in the U.S.A. In space the initial studies of ESA and NASA led to the selection of the SOHO space project in 1988. The SOHO payload included the Global Oscillations at Low Frequency (GOLF, Gabriel et al. 1995), the Michelson Doppler Imager for Solar Oscillation Investigation (MDI/SOI), and the Variability of Irradiance and Gravity Oscillations (VIRGO) instruments. As stated by Appourchaux \& Pall\'e~(2013), in these days, the detection of g modes was thought to only be a matter of time. In the end, it still required a further 20 years of effort. Some possible results were published during these 20 years using GOLF data. One $l=1$ mixed p-g mode was reported at 284.7~$\mu$Hz (Gabriel et al. 2002), and several candidates were suggested in the vicinity of $220\ \mu$Hz (Turck-Chi\`eze et al. 2004). Garc\'ia et al.~(2007) reported the possible collective detection of a g-mode signature in a lower frequency range between 25 and 140~$\mu$Hz, with a possibly rapid rotation of the solar core up to five times the
rotation of the remaining radiative zone. Possible detections of five $l=1$ g modes were reported by Garc\'ia et al.~(2011) in the frequency range between 60 and $140\ \mu$Hz, with rotational splittings in the range 850 to 950~nHz consistent with a fast sidereal rotation (1760 to 1960~nHz) of the g modes themselves, measured using the rotation-corrected, m-averaged spectrum technique from Salabert et al.~(2009). An extensive review of different techniques of detection, as well as possible results, can be found in Appourchaux et al.~(2010).  So far, none of these possible results has been confirmed.

\section{Looking differently and elsewhere}

The  search for g modes must overcome the difficulty that the
modes are trapped in the deepest part of the solar body and are evanescent in the convective layers, so that their amplitudes are not easily detectable at the solar surface.  The p-mode amplitudes themselves decrease dramatically with decreasing frequency (left side of Fig.~1), falling below the increasing background noise level before reaching the frequency range around the fundamental  p-mode oscillation, where a few mixed modes (driven by both compressibility and buoyancy) are thought to exist. In this range, roughly speaking between 0.25 and 0.30~mHz, the lowest frequency p modes and highest frequency g modes are expected to exist with similar amplitudes at the solar surface, an amplitude of the order of 1~mm\,s$^{-1}$ with a large uncertainty on any prediction (see again Appourchaux et al.~(2010) for a comprehensive review), and they have not been detected so far. The amplitude uncertainty remains very high in the g-mode domain, especially at lower frequencies, where the background noise increases. It has never been demonstrated that the signal-to-noise ratio (S/N) at these lower frequencies can improve in any way. This is the fundamental reason why purported detections of g modes in this region of the spectrum have always been doubted and have never been confirmed, despite numerous efforts.

Kennedy et al.~(1993) suggested that the g modes might be detected as a source of modulation of p-mode frequencies, which would manifest themselves as sidelobes in the p-mode frequency profiles. However, as noted by the authors, convective noise is a problem. The sidelobes cannot be narrower than the p-mode lines, and  the p modes are more sensitive to turbulent motions near the surface than they are to the g-mode motions in the deepest solar structure. Despite serious efforts in this direction, this approach has not been successful so far.  Another approach was also suggested by Duvall~(2004): modifications of time-distance helioseismology to enable travel time measurements for short time intervals. This has not been successful either.

 Although based on the same idea of time distance measurement,  our approach is nonetheless different.  We seek to identify a differential parameter in the p-mode frequency spectrum that can significantly decrease the sensitivity to this convective turbulence. Such a parameter should ideally  exploit the p-mode frequencies measured by the GOLF instrument, taking advantage of the fact that these full-disk p modes are modes of the lowest degrees, that is, from 0 to 3, which propagate through the solar core.

The apparent ideal candidate would be the so-called small separation, that is, the frequency separation between p modes of degrees 0 and 2. This was immediately identified, after being first resolved by the South Pole data set in 1980 (Grec et al. 1980, 1983), as a parameter probing the solar core because it measures a difference of two kernels that are essentially identical in the upper layers and different in the deepest layers. Unfortunately, it requires more than two days of measurement to be resolved, so that only g modes with periods longer than these two days could be tracked. These very long period g modes are not accessible in practice, largely because of the high density of g modes per frequency bin.

The second-best choice is then the so-called large separation, or rather half this large separation  (between even 0-2 and odd 1-3 pairs of p modes), which defines the quasi-vertical lines in the so-called Echelle diagram.  
The large separation is linked to the propagation of the acoustic waves through the Sun and is proportional to the mean density of the star. Thus, any modulation of the solar structure produces a modulation of the large separation. If these modulations are produced by the action of g modes on the geometry or density of the deep solar core, they should present the same spectral properties as those expected for the g modes. We can also seek for g-mode signatures in the large separation signal.
The large separation offers an additional  benefit of a lower sensitivity to surface effects when adequate care is taken to measure this separation. A detailed study of the Echelle diagram shows that the almost vertical lines have a general S-shape that contains the signature of the depth of the helium ionization layer. This layer, not far below the solar surface, is again a turbulent region, which we wish to avoid. We therefore have to limit the frequency range used for measuring this large separation to the straightest line part of this Echelle diagram, in practice, to the frequency range  between  2.32 and 3.74~mHz, which is
defined precisely to ensure that it contains an integer number of equidistances. 

Even in this limited range, the frequencies plotted in the Echelle diagram still display small oscillations around the straight lines that contain the signature of the depth of the convective zone. This boundary layer, named the tachocline, is another dynamic region that we also wish to avoid here. It is then essential to use the broadest possible frequency range in measuring the large separation to reduce the sensitivity of its fluctuations to this tachocline region. The range 2.32 - 3.74~mHz appears to be the best compromise between the need for limiting the range to the most  linear part and the need for the broadest possible range. 

There are two more reasons to justify the choice of this frequency range limitation. Below 2.3~mHz, the individual spectra are not sufficiently sensitive, and the p-mode peaks are lost in background noise with 8 hours of integration (see below). At the other end, above 3.7~mHz, there is a sharp identified increase of the p-mode parameter sensitivity to all solar surface effects (Salabert et al. 2002).  Frequency and line-width changes with the solar cycle are much larger than in the lower p-mode frequency domain, which  would introduce more undesired errors in the search for the subtle modulations produced in the solar core.


We therefore use this semi-large separation, measured between 2.32 and 3.74~mHz,  as a tool for the g-mode search. A p-mode spectrum obtained from an 8-hour time series reveals the discrete structure of its frequency distribution, characterized by this pseudo-periodicity on the order of 67.5~$\mu$Hz. This spacing corresponds to the inverse of the round trip time $T$ of an acoustic wave traveling through the solar center. One interesting consequence is that the full-disk velocity signal has a strong correlation peak at a lag $T$ of about 4 hours and 6 minutes (which is on the order of the inverse of 67.5~$\mu$Hz). This was used to advantage by the IRIS network team (Fossat et al. 1999) in an optimal gap-filling technique. In the present context, we consider $T$ as the optimal parameter to be monitored in the search for its possible modulations produced by a change of structure located inside this core. Because we cannot filter the frequency spectrum before sampling the time series $T(t),$ we need to be careful about aliasing, which occurs when the spectrum is folded about the Nyquist frequency. An easy and efficient technique consists of sampling the signal twice faster than the integration time, which means a 4-hour sampling in the present case. The Nyquist frequency is then doubled from 1/16h = 17.36~$\mu$Hz to  34.72~$\mu$Hz. This avoids the spectrum folding, which would be around 17.36~$\mu$Hz if we were sampling at 8-hour intervals. The search for g-mode frequencies is then possible in the low-frequency range that corresponds  to periods longer than 8 hours, even if the part of the spectrum approaching 34.72~$\mu$Hz (inverse of 8 hours) is progressively filtered and contains aliases of frequencies coming from beyond 34.72~$\mu$Hz. It still provides access to potential g-mode frequencies in this range.

\begin{figure}
\centering
\includegraphics[width=\columnwidth]{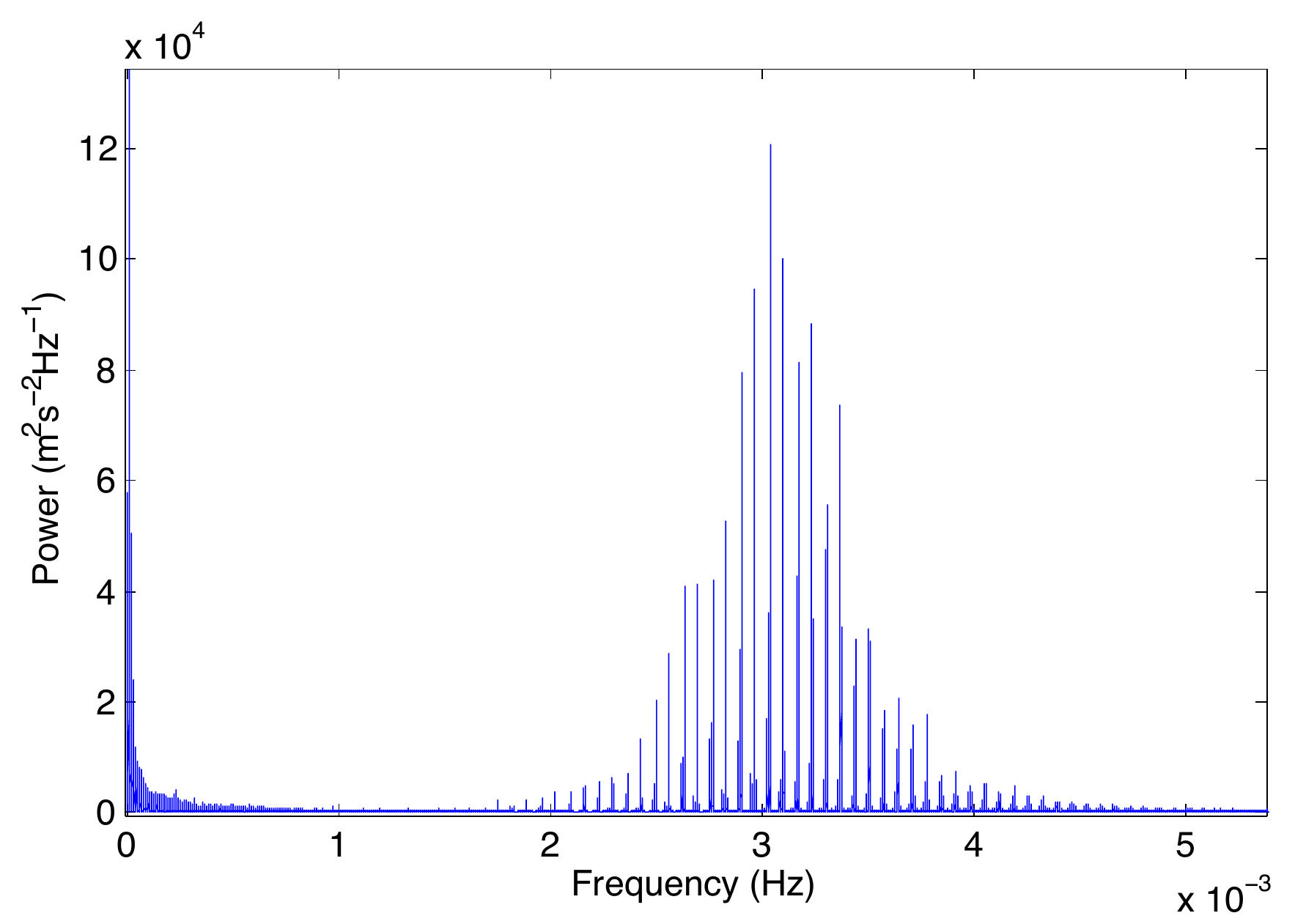}
\caption{Power spectrum of 16.5 years of full-disk velocity recorded by the GOLF instrument on board the SOHO spacecraft.}
\label{fig:PS16y}
\end{figure}

We worked with a GOLF time-series calibrated in velocities as explained in Garc\'ia et al.~(2005), which starts at 0:00:30 (T.A.I.) on April 11, 1996. It is sampled at 80~s. For the 16.5 year interval with 8-hour segments sampled every 4 hours, we start with 36131 possible samples, but because of data gaps (especially the so-called SOHO vacation in 1998), we finally have 34612 8-hour p-mode spectra.

The general process developed for obtaining the time series $T(t)$ can be summarized as follows:

1. Split the GOLF  time series into 8-hour long sub-time series at 4-hour intervals.

2. Compute the power spectrum of each 8-hour long time series oversampled by filling zeroes up to $10^6$ s to obtain a convenient 1~$\mu$Hz frequency bin (the mean of all these spectra is shown in Fig.~2).

3. Divide each power spectrum by a Gaussian function ($\exp(-x^2/2\sigma^2)$) centered at 3.22~mHz with $\sigma=0.39$~mHz, as an approximation of the envelope of the mean spectrum, and select  the [2.32 - 3.74] frequency range. An example of such a band-limited envelope-corrected spectrum is shown in Fig.~3.

4. Compute the power spectrum of each of these windowed power spectra oversampled by filling zeroes up to 125~mHz to obtain a time bin of 8 s. These power spectra of power spectra result in a function that has time as its abscissa, with their main peak at this period $T$ of approximately 4 hours, as presented
in Fig.~4, which shows one example of these power spectra, or in Fig.~5, which is the sum of 34612 similar spectra.

5. Measure the precise location of the centroid of the main peak around its mean value of 14807~s using a second-order polynomial fit.

We now describe these steps in some more detail.

The nearly 67.5~$\mu$Hz equidistance is measured by a Fourier analysis that requires two levels of oversampling.  A simple fast Fourier transform (FFT) power spectrum of power spectrum would return the initial sampling time of 80~s, which  is not fine enough for our purpose. The spectra shown in Figs.~2 and 3 are oversampled to a convenient bin of 1~$\mu$Hz (step 2), and then the 34612 spectra of spectra (see Figs.~4 and 5) are also oversampled to a bin of 8~s, which was chosen to be smaller than the r.m.s. scatter of $T$ (step 4). We show that these oversampling steps do not affect the statistical properties of the final product of this step of analysis, that is, the power spectrum of $T$ (Fig.~7).

For the most precise estimation of the equidistance through this Fourier method, it is also necessary to ensure that the entire selected frequency range contributes and not just the few highest p-mode peaks around 3.2~mHz. Each of these individual frequency spectra is then "flattened" (step 3) by dividing it by a Gaussian function (centered at 3.22~mHz with $\sigma= 0.39$~mHz) that approximates the envelope of the mean (16.5 year) spectrum (Fig.~2), thus allowing optimal use to be made of the resolution provided by the 1.42~mHz bandwidth. Figure~3 is an example of such a selected and flattened individual spectrum.  
Each of the 34612 individual envelope-corrected 8-hour p-mode spectra is then analyzed by Fourier transformation, from which its respective mean p-mode separation is determined.
This fifth step of the analysis is not made through a fit of the GOLF signal autocorrelation peak around 4 hours because that would not permit the correction of the power envelope, which is an essential step of this analysis.

We note that our spectra such as shown in Fig.~4 are computed from spectra such as shown in Fig.~3, minus their mean value, and translated at abscissa zero. This means that the first point of Fig.~4 is at zero and suppresses the 5-minute modulation,
which would remain inside the 4-hour peak, whose position can then be measured more precisely. These spectra are arbitrarily normalized at the height of the main peak, since each spectrum is no longer exactly an autocorrelation that would automatically be normalized at 1 on the zero lag. Only the precise position of the 4-hour peak centroid (our parameter $T$) is relevant in our analysis, and our general guideline is then to minimize the amplitude of the random fluctuations of this parameter $T$ around its mean value.  

  Each value of $T$ is estimated by a second-order polynomial fit around the peak profile (step 5). The choice of the width over which this fit is computed is again guided by the goal of minimizing the random fluctuations of T. A priori it should be made at least on the true bin size defined by the inverse of the range from 2.32 to 3.74~mHz, that is, about 700~s. In practice, the polynomial fit  made on a range of $\pm 800$~s  around 14800~s minimizes the scatter of $T$.  This is on the order of the full width of this peak and not at all the best fit on the peak profile, but it is the least noisy estimate of the peak centroid. The profile of this peak could be sensitive to the way the raw intensities are corrected and then calibrated into velocities, which is indeed the case for the amplitude of the peak. However, the centroid of the peak is our relevant parameter
here, and there is no reason for it to be calibration dependent, either through physics assumptions or mathematical processes. This is indeed one of the reasons why we selected this parameter in our g-mode search.  The peak fit sometimes fails; a limit of 240~s was therefore set to the values of the fluctuations of $T$ to eliminate these unrealistic poor-fit results. Less
than 1 percent of the values is rejected, so that the number of non-zero values of $T$ is finally 34261. The effects are all negligible for clock
and timing. The SOHO on board timer is always maintained within 20~ms of TAI. We checked that the few shifts that occurred in the GOLF sampling during the operations had no signature in all of the processes we describe here. We also checked that neglecting the SOHO-Sun distance variation through the orbit has no influence on our results.
\begin{figure}
\centering
\includegraphics[width=\columnwidth]{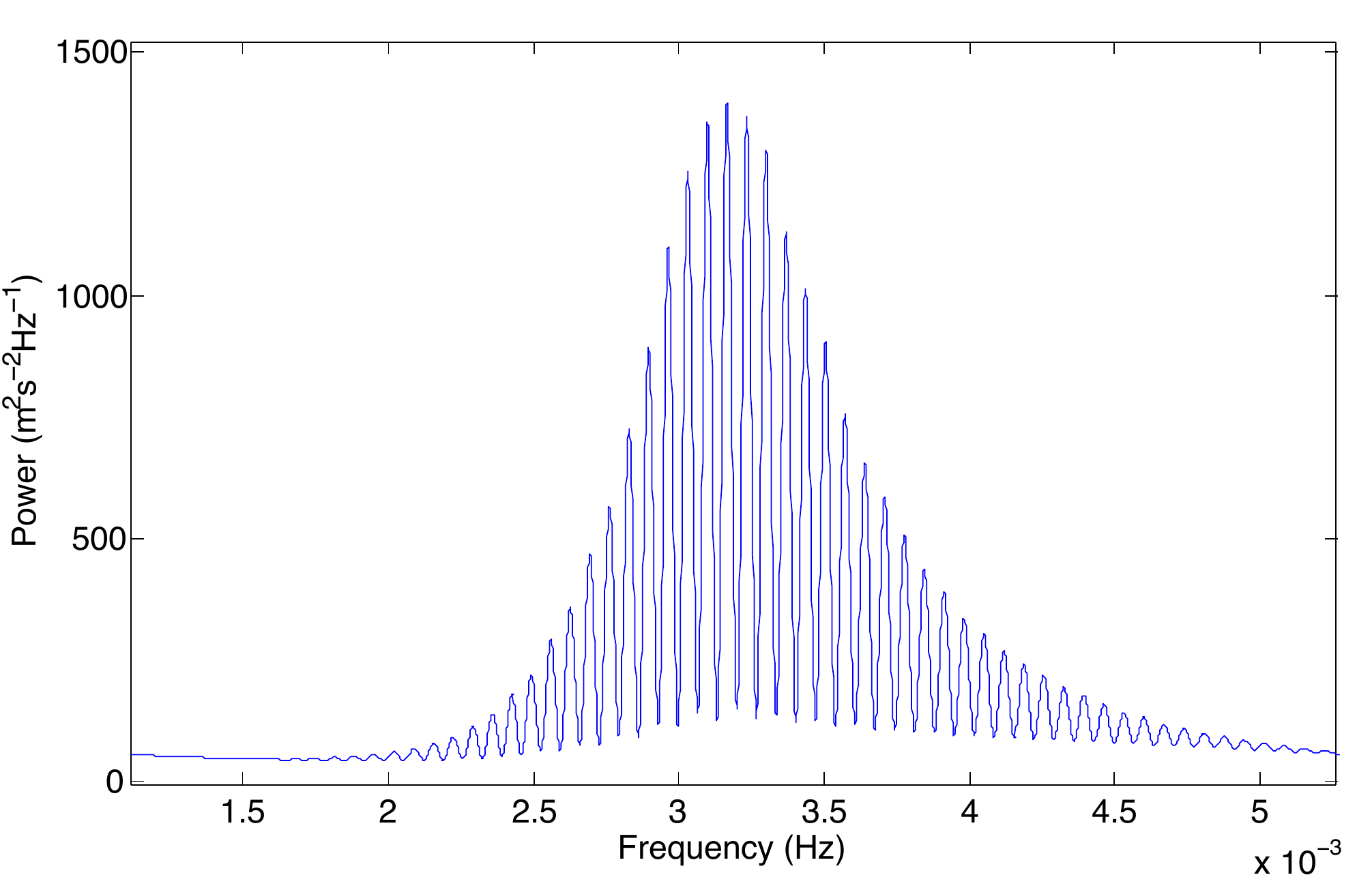}
\caption{From the same 16.5-year data set as Fig.~1, this is an average of 34612 spectra computed from 8-hour selections taken at intervals of 4 hours.}
\label{fig:av_spec}
\end{figure} 

\begin{figure}
\centering
\includegraphics[width=\columnwidth]{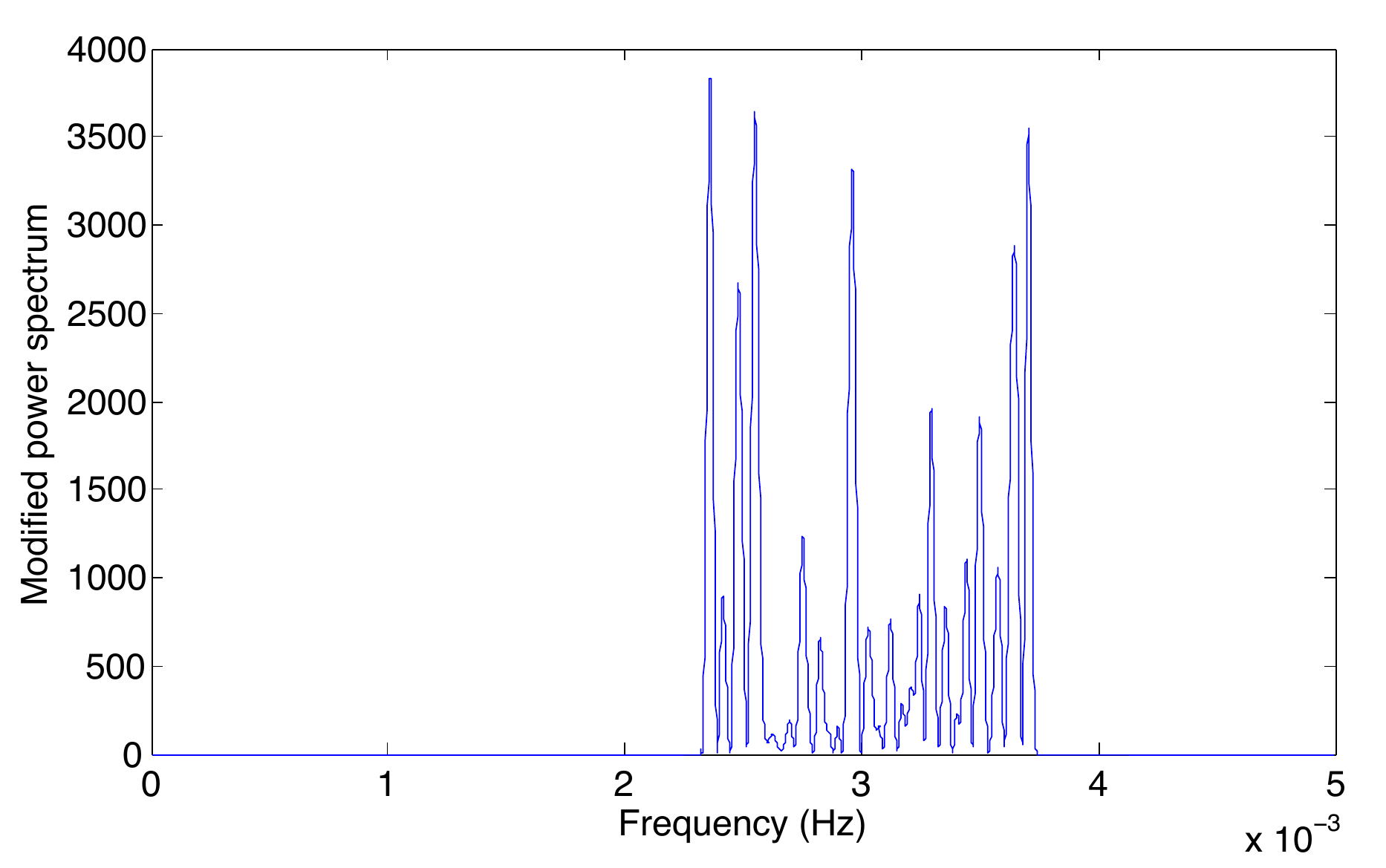}
\caption{Example of one of the 34612 GOLF power spectra, limited to the range 2.32 - 3.74~mHz and divided by the envelope of the mean spectrum of Fig.~2.}
\label{fig:ps_example}
\end{figure} 

\begin{figure}
\centering
\includegraphics[width=\columnwidth]{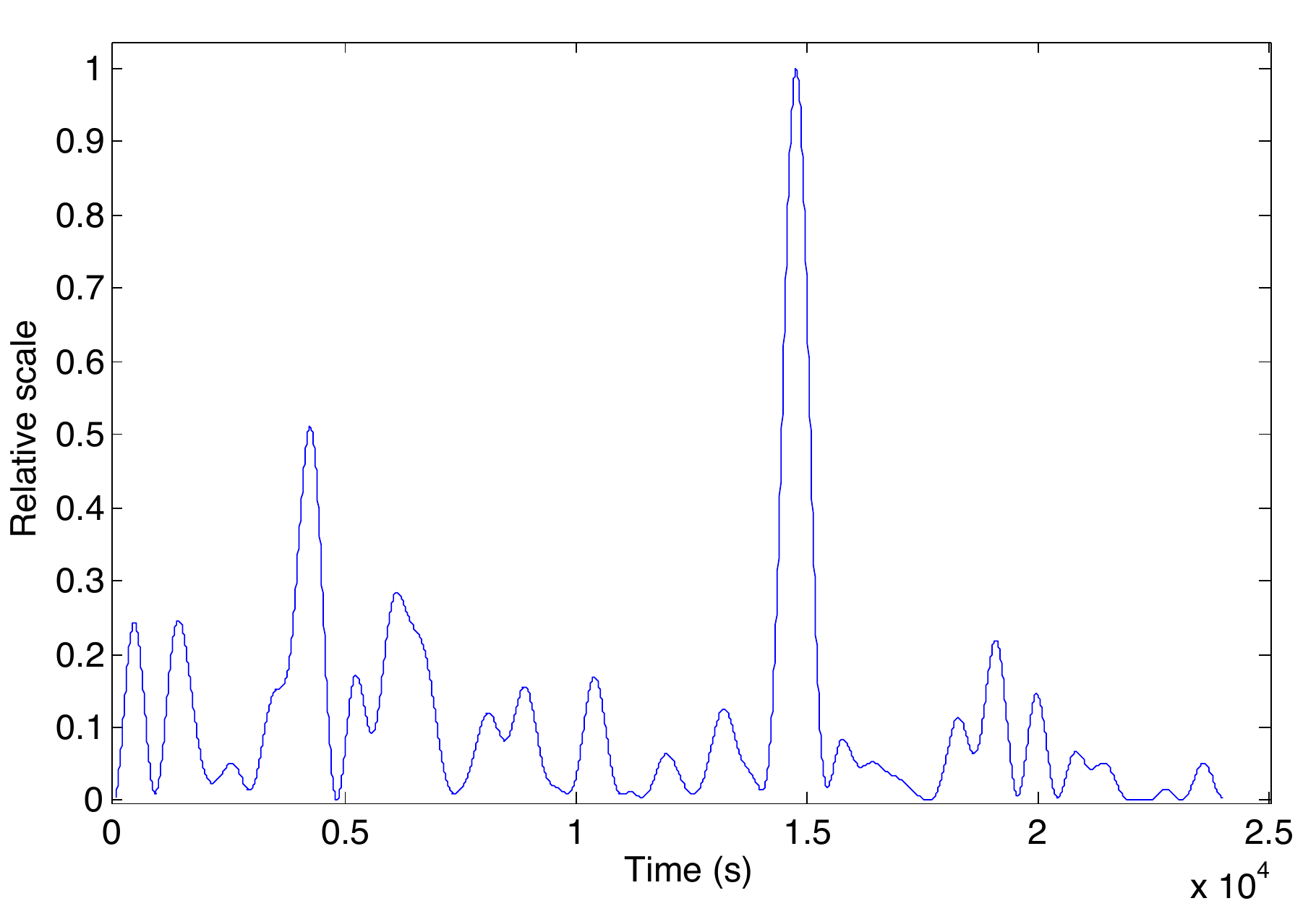}
\caption{Power spectrum of the spectrum shown in Fig.~3. The abscissa is in seconds and shows a main peak at a time $T$ of about 14800~s, which is the round trip travel time of an acoustic wave through the solar center. }
\label{fig:ps_of_ps}
\end{figure} 

\begin{figure}
\centering
\includegraphics[width=\columnwidth]{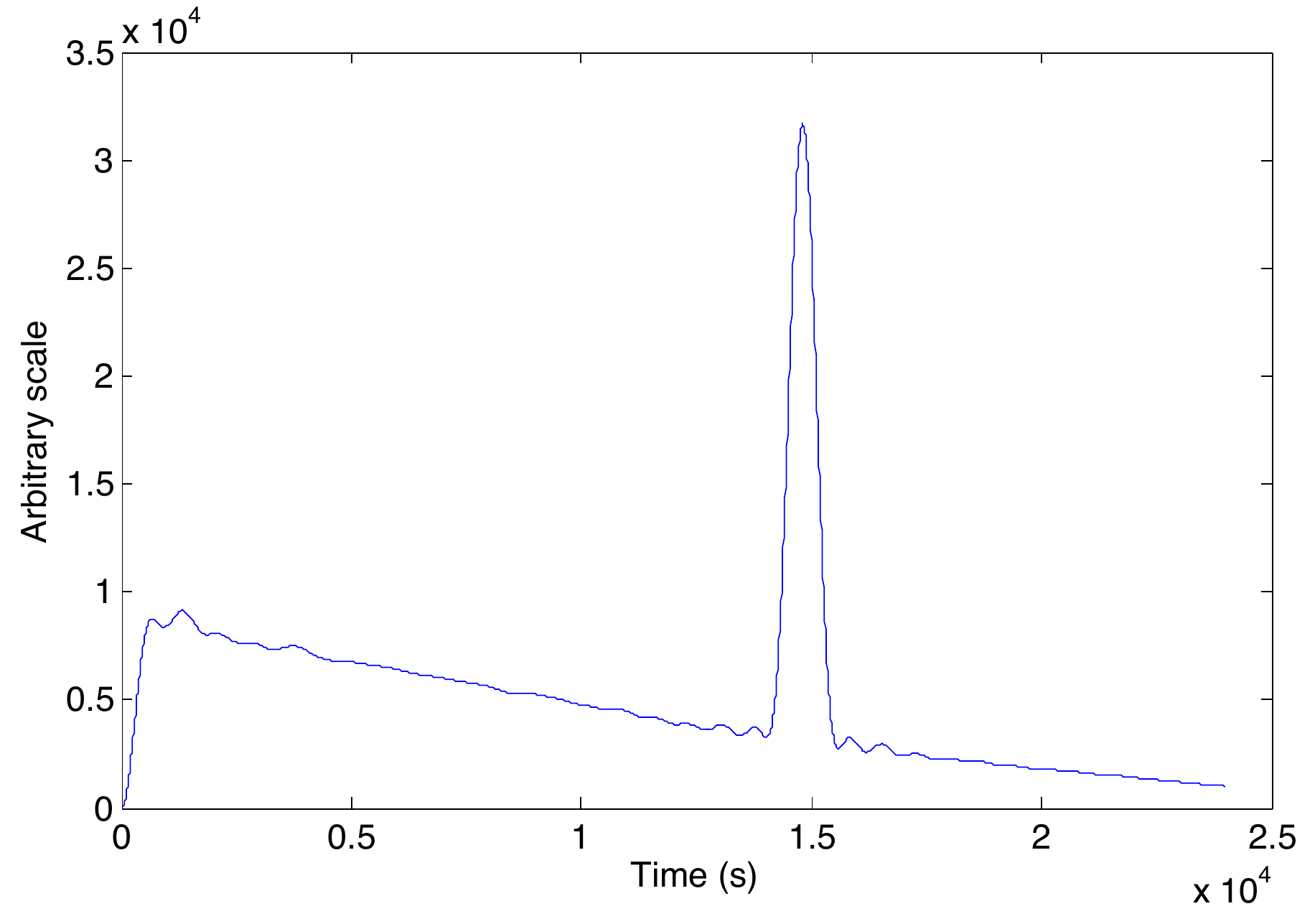}
\caption{Sum of 34612 power spectra similar to the spectrum shown in Fig.~4.}
\label{fig:sum_of_ps}
\end{figure}

\begin{figure}
\centering
\includegraphics[width=\columnwidth]{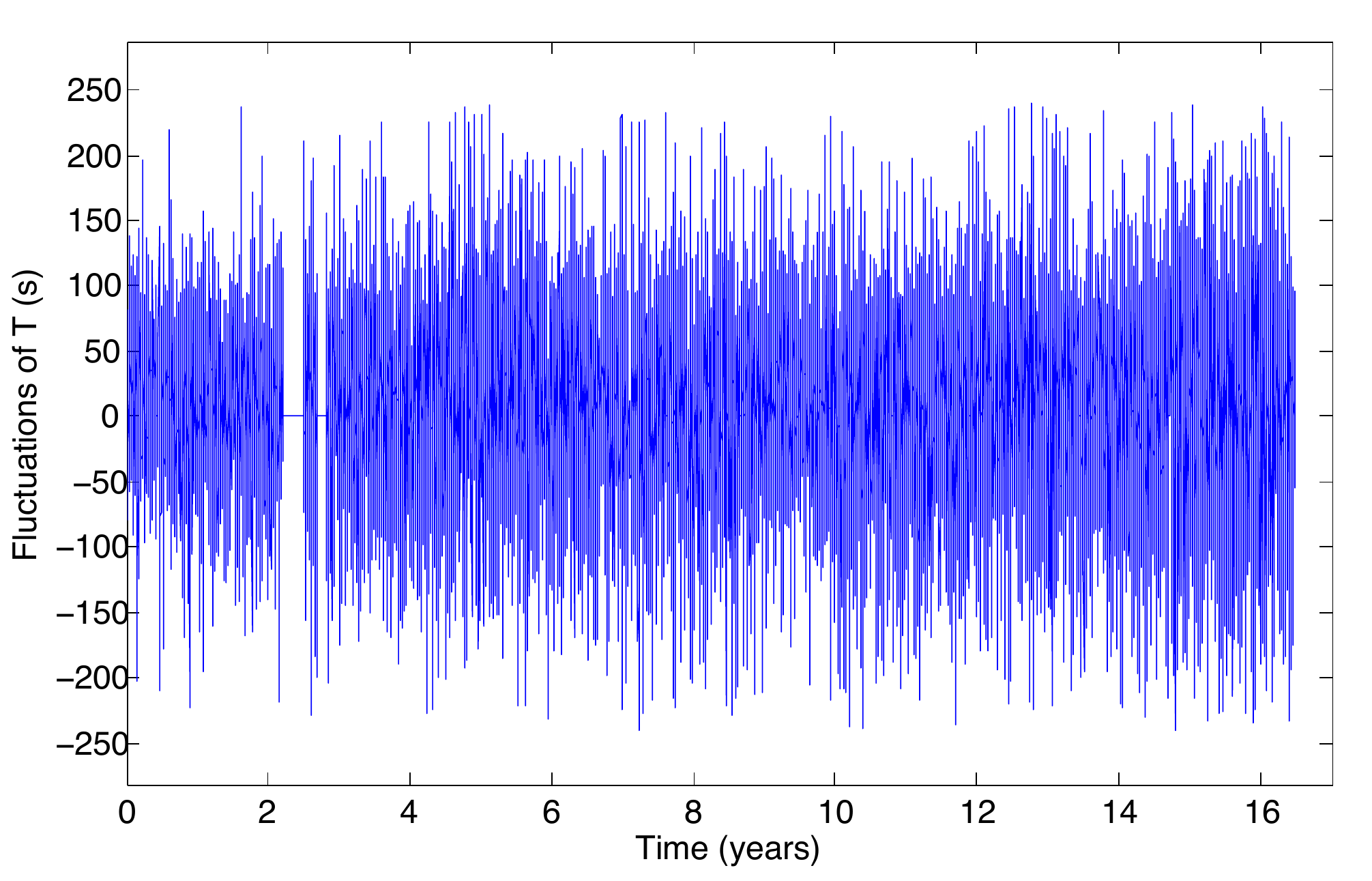}
\caption{16.5-year time series of 36131 values of the time $T$ fluctuations defined in Fig.~4. The number of non-zero values is 34261. }
\label{fig:T}
\end{figure} 

Figure~6 shows the time series made up from 36131 consecutive values of $T_i$ fluctuations (where $1\le i \le 36131$, including gaps). This is the new 'signal' to be analyzed in the hunt for g modes.  It is interesting to know that the mean value of $T$, when it is measured as  we did, is 14807~s, or 4 hours, 6~min, 47~s. This can be seen as the mean return travel time of a sound wave from the visible surface through a diameter. The measured r.m.s. value of the scatter of $T$ around this mean value is 52 seconds. The distribution around the mean value is nearly Gaussian with this $\sigma$ of 52~s, with slightly extended tails on both sides that correspond to the poor fits and were cut, as mentioned above, at $\pm 240$~s.

We checked that our parameter $T$ is remarkably insensitive to surface effects such as solar activity or orbital velocity residuals (again, it was carefully defined for that purpose). For instance, Fig.~6 just marginally shows a slight increase of noise amplitude with time, while the GOLF signal itself shows it more substantially, as well as showing some consequences of a few technical changes on the instrument through these years that are totally invisible here. $T$ is also  insensitive to the detailed characteristics of the calibration steps used to process the GOLF signal, which will in any case reveal the p-mode spectrum in our selected range. 

Figure~7 shows the resulting power spectrum of the 16.5-year time series of $T$ fluctuations. Although, as discussed above, this spectrum is limited to an extremely low frequency domain (below $34.7\ \mu$Hz, i.e., the lower 0.5 percent of the bandwidth shown in Fig.~1), it is remarkably flat, whereas the GOLF signal spectrum itself is characterized by a very strong $1/f^2$ dependence in this range (Fig.~1), where it would not make sense to look for g modes. In addition, Fig.~8 shows an enlargement on the extreme lowest frequency part  that illustrates the absence of signature of either the solar cycle period or the orbital period of one year, again demonstrating the very low sensitivity of $T$ to the surface velocity.  

\begin{figure}
\centering
\includegraphics[width=\columnwidth]{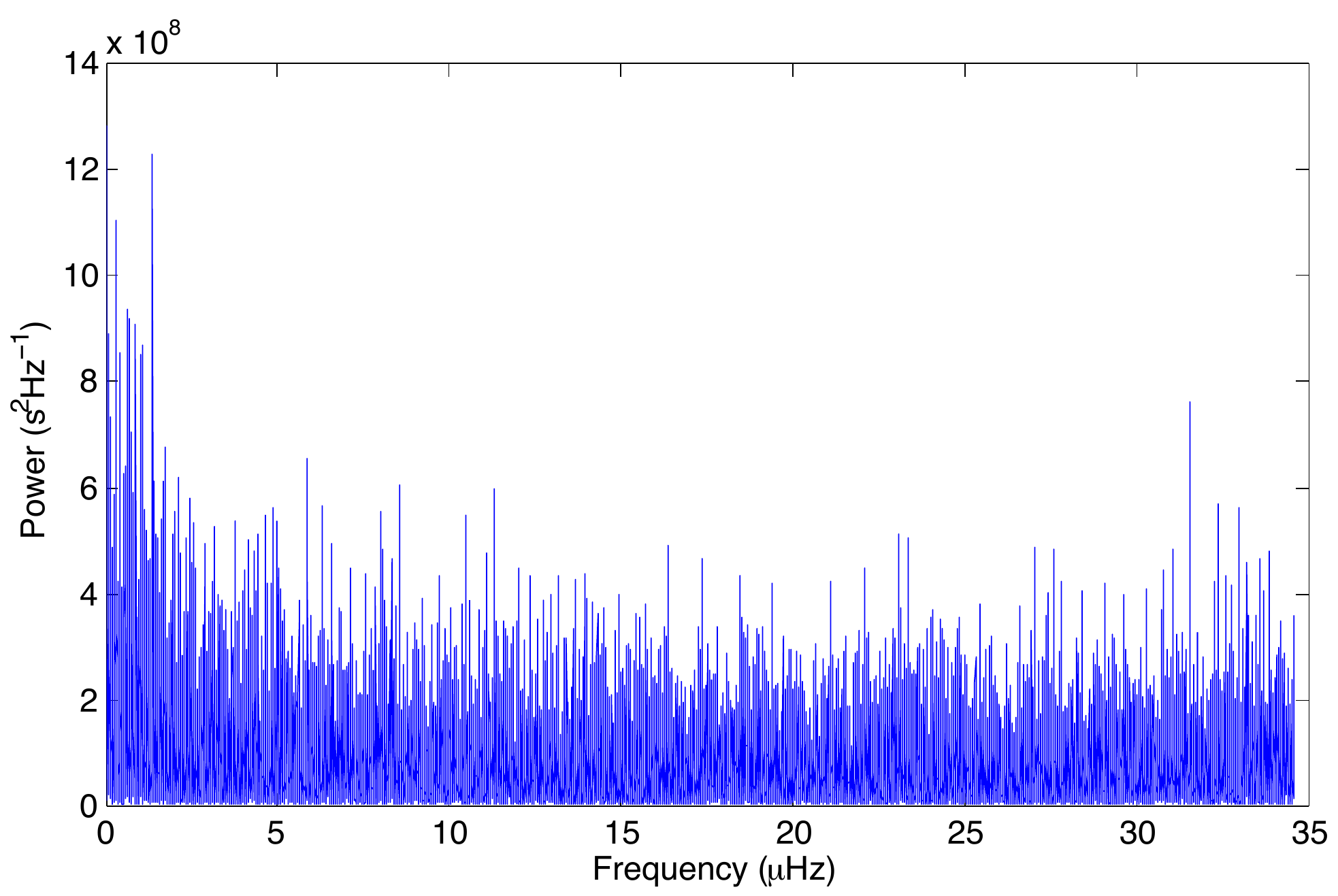}
\caption{Power spectrum $P_\mathrm{S}(\nu)$ of the time series shown in Fig.~6.  }
\label{fig:PS}
\end{figure}

\begin{figure}
\centering
\includegraphics[width=\columnwidth]{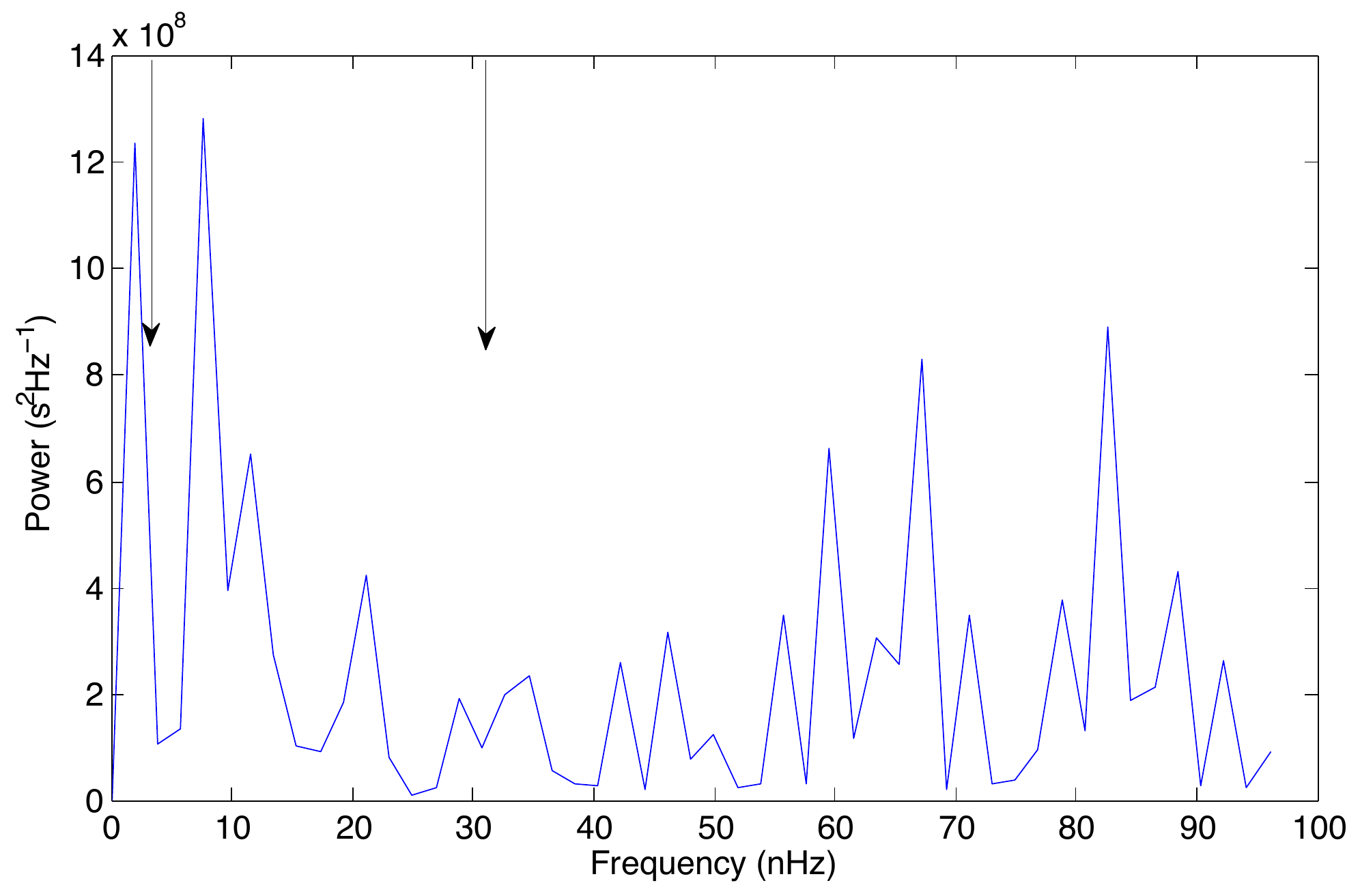}
\caption{Enlargement of the lowest frequency range of Fig.~7. The arrows indicate the frequencies of the solar cycle and of the one-year orbit.  }
\label{fig:PS_zoom}
\end{figure} 

We now use this power spectrum  to determine low-frequency asymptotic values of g-mode parameters, taking the very low frequency range into account over which this asymptotic search is best possible.    In this analysis, the relevant asymptotic g-mode properties for a given degree are on one hand the equidistant period spacing between consecutive radial harmonics, and on the other hand, assuming a uniformly rotating core, a constant frequency splitting value for the triplet structure of the $l=1$ dipolar modes and for the quintuplet structure of the $l=2$ quadrupolar modes. We show below that only degrees 1 and 2 have indeed been detected.

\section{Real GOLF data versus artificial data sets}

At various steps of the analysis below, it will be important to check that the analysis itself is not responsible for creating the results that we will claim. For this reason, it is essential to compare the real GOLF data with artificial datasets presenting all the GOLF p-mode amplitudes and noise characteristics and for which the temporal coverage is identical to the GOLF window function.  

In order to create a simulated time series with the same spectral content and the same statistical behavior as the actual GOLF data, we use the fact that the p modes are damped and randomly excited so that the data can be regarded as a stationary random process. 
Therefore, the periodogram of the time series, calculated as the square modulus of the FFT of the time series, can be regarded as a particular realization of the power spectrum density (PSD) of the solar signal. The PSD is the asymptotic limit of the Fourier transform of the autocorrelation of the solar signal observed for an infinite duration. When the data are evenly spaced, the FFT calculated on the same number of points as the time series for a finite observation is producing independent values, bin by bin, for both the real an imaginary parts, which  are
also normally distributed, so that its square modulus follows a $\chi^2$ distribution. It is then possible to produce simulated time series $y_\mathrm{sim}$ with the same spectral contents and statistical properties by multiplying the square root of the PSD of the data by two random normal distributions to produce the real and imaginary parts of the Fourier transform of the time series, in the form

\begin{equation}\label{eq:simu}
y_\mathrm{sim} = FFT^{-1}(X)
,\end{equation}
with
\begin{eqnarray}
X(\nu_i)&=&\sqrt{PSD}(N_1+\mathrm{j}N_2),\ \  0<\nu_i\\
X(-\nu_i)&=&X(\nu_i)^{\star}
,\end{eqnarray}
where $N_1$ and $N_2$ are two normally distributed random series with half the number of points of the GOLF full time series. This method has been extensively studied by Percival~(1992) and has been used in helioseismic data analysis (see, for instance, Fierry-Fraillon et al. 1998; Appourchaux et al. 1998).

In order to estimate the PSD of the GOLF data, we used the method developed by Welch~(1967). We calculated 120 periodograms corresponding to successive 90-day periods of the total 16.5-year time series overlapped by 50\% and averaged them. The resulting spectrum was convolved by a triangular $0.1\ \mu$Hz window function to obtain a smoother estimate of the PSD. Figure~9 shows the square root of the PSD superimposed on the GOLF time series FFT modulus. The PSD estimate contains all the known properties of the p-modes spectrum, as its resolution is sufficient to resolve all the components of the split p modes. Therefore, any combination of split modes that could propagate through any subsequent step of our analysis would exist in the simulated data. In contrast, the possible modulation of the p-mode separation by g modes is completely suppressed in the averaging. It should be noted that the simulated data not only have the same  spectral contents for the p modes, including the presence of all modes up to $l=4$, with correct profile widths, amplitudes and asymmetries, but also quantitatively reproduce the surrounding noise level.

\begin{figure}
\label{fig:AmpSpec}
\centering
\includegraphics[width=\columnwidth]{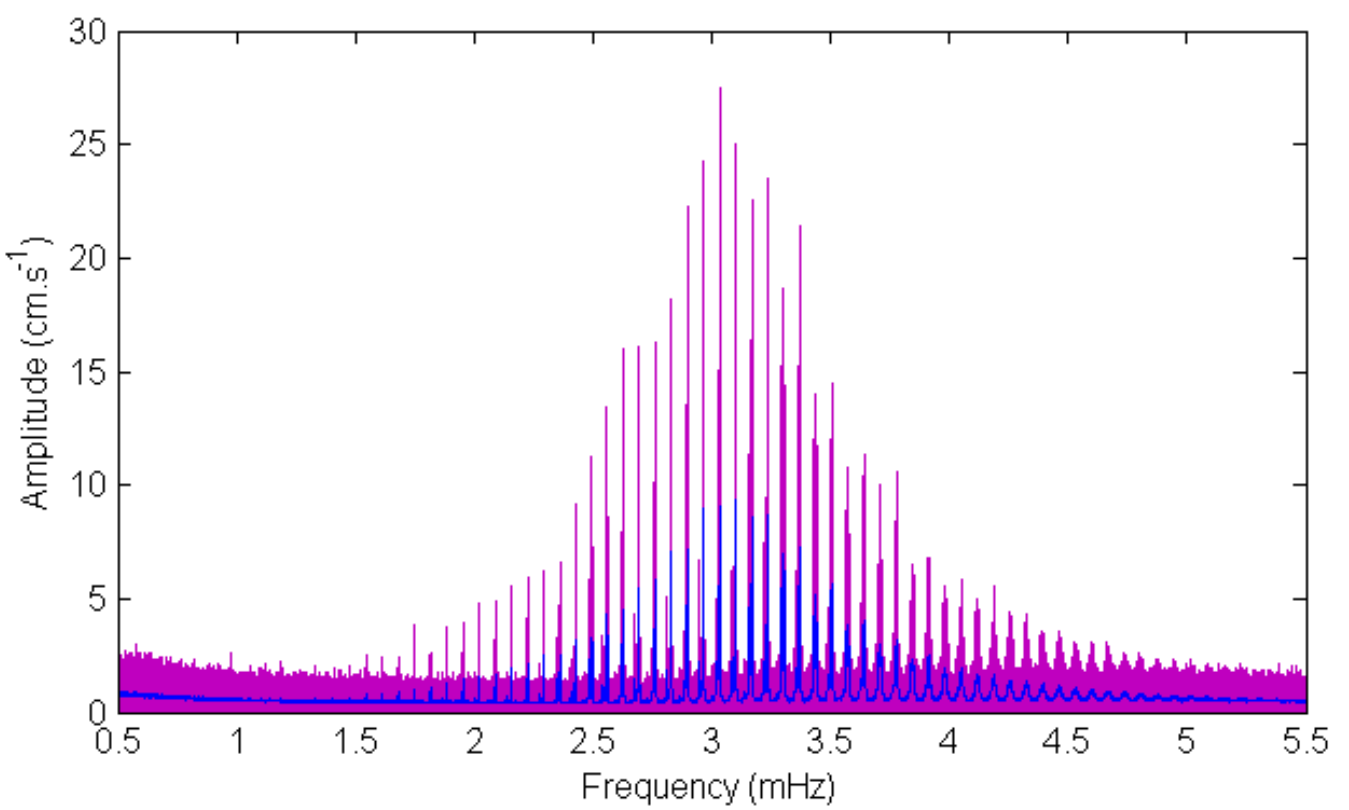}
\caption{Estimate of the amplitude spectrum (square root of the density power spectrum) in blue, superposed on the modulus of the Fourier transform of the GOLF time series (in magenta). This estimate was used to calculate simulated time series.}
\end{figure} 

This process can be compared to the method described by Percival~(1992). We added one specific step to take the GOLF data set specificity
into account, however. The GOLF data duty cycle is close to 96\% in this 16.5-year time series. A few gaps exist, in particular at the beginning of the time series (the SOHO vacation). Some variations of the mean amplitude can also be seen during the 16 years of observations, corresponding to the change from one side of the sodium lines to the other, as well as a yearly variation linked to the  orbital ellipticity. These features would result in an imperfect independence between the frequency bins in the FFT. This was taken into account in the simulated data.  For that purpose, the time series $y_\mathrm{sim}$ were multiplied by the total GOLF window function and by an amplitude modulation calculated from the GOLF dataset with a 3-month moving average. By doing so, we ensure that the simulated spectra have the same statistical behavior as the original time series. It is then possible to analyze the simulated spectra with each step of the procedure that was applied to the original GOLF data.

Ten such artificial GOLF time series were produced. They were used for the first step of our analysis detailed so far:  the production of the time series of $T(t)$. The r.m.s. values of the fluctuations of these ten time series lie between 49.5 and 51~s, all close to but slightly lower than the 52~s obtained for the GOLF data set.

\section{Rotational splitting and solar core rotation}

Very many modes of degrees 1 and 2 are available that have detectable amplitudes in the very low  frequency range we are considering.  The best first step in this case, as suggested many years ago (Fossat et al. 1988),  is to look for the rotational splitting that is assumed to be approximately independent of the radial order for a given degree of the g modes in the asymptotic approximation. The obvious approach involves computing the autocorrelation of the power spectrum shown in Fig.~7. To maximize the statistical improvement, we used the broadest possible frequency range, just avoiding the very lowest part, where a slight increase in noise is visible, and where we also know that the g-mode density per unit frequency (if this spectrum contains any g-mode signature, of course) becomes too crowded to permit any possible use. We also used a version of this spectrum smoothed over 6 bins, which allows contributions from peaks with frequencies that deviate slightly from the strict asymptotic equality of all individual splittings. Figure~10 shows this autocorrelation computed over the  frequency  range between 5.5~$\mu$Hz and the Nyquist frequency of 34.7~$\mu$Hz.  About 100 modes of degree 1 and 170 modes of degree 2 could exist in such a broad range.

The main peak of this autocorrelation, around  210~nHz, is highly significant for two reasons. First, when we use the standard deviation $\sigma$ of the autocorrelation, which has a Gaussian distribution, the peak at 210~nHz is at 4.7~$\sigma$ above the mean level of the autocorrelation. With the 6-bin smoothing and a Gaussian distribution of 2000 bins of this autocorrelation in the range displayed in Fig.~10, that is, around 300 independent values, this implies a low probability of lower than $10^{-3}$  for this peak to be produced by a random fluctuation. Second, this peak remains the main peak in various attempts to reduce the statistics, either by exploring a narrower frequency range, or by reducing the duration of the 16.5-year initial data set to about half. Figures~11 and 12 show the autocorrelation computed on the two independent frequency ranges, from 5.5 to 13.2~$\mu$Hz in Fig.~11, and from 13.2 to 34.7~$\mu$Hz in Fig.~12. This shows
that this correlation peak is clearly present over a very broad frequency range.
This is a first indication that many asymptotic g-mode signatures are present in this broad range, and this is confirmed in the next section by means of the other asymptotic parameter, namely the period equidistance.
 Figure~13 shows the increase in 210~nHz peak significance with increasing the duration of the time series. The length of the time series can be adjusted by adding data points to either the beginning or the end, and this figure is the average of the two possibilities. This figure shows that the autocorrelation clearly becomes significant with about 8 to 9 years of data. 

The ten sets of artificial data never produce an autocorrelation value above 4 $\sigma$ through the
same analysis, and never produce a highest peak at 210~nHz. This excludes the idea that the peak could be an artifact produced by a subtle effect of the p-mode splittings themselves.  As an example of these autocorrelations from artificial data sets, Fig.~14 does not show any outstanding peak, and the highest peak, not at 210~nHz,  stands at 3.3~$\sigma$. As we mentioned, interpreting
this as a g-mode splitting signature assumes a constant splitting value of the asymptotic approximation, and this is probably true only within 3 percent in the highest part of our frequency range (Berthomieu \& Provost 1991). It can be verified that the smoothing of the power spectrum improves the S/N of this correlation peak. The maximum benefit is obtained by a smoothing on 6 to 8 bins, which is consistent with this small departure from the asymptotic approximation.  If the same autocorrelation is computed without the 6-bin smoothing, the main peak at 210~nHz is spread across 6 bins, and we can note that bin number 2 at an abscissa of 1.92~nHz is at lower than 0.02, which proves that there is essentially no correlation between consecutive bins in the power spectrum.

Figure~15 shows the histogram of the ten autocorrelations of the same frequency range of the power spectra of artificial time series, all taken together and individually centered around their mean value (which is not exactly zero because of the 6-bin smoothing and the slow variations of the background noise). The r.m.s. value is 0.0166 and the highest of these 20000 values is at 3.8~$\sigma$.  This indeed corresponds to a probability of 1/20000 in a normal gaussian distribution. The 210~nHz peak of Fig.~10 (where the r.m.s. value is also 0.0166, which confirms the internal consistency of the simulation) that stands at 4.7~$\sigma$, has a probability of $2\times10^{-6}$ to be a random outcome. These numbers are consistent with the estimated probability of lower than $10^{-3}$ mentioned above that an autocorrelation of 300 independent values contains one of these as high as this main peak, as $300\times 2\times 10^{-6} = 0.6\times 10^{-3}$. Its position is shown by an arrow in Fig.~15.

The statistical  significance of this correlation peak is then quite strong, and this test on artificial data demonstrates that it is not produced by a subtle effect of the p-mode splittings. The subsequent analysis has proved that it can really be interpreted as the signature of g-mode triplets. 

\begin{figure}
\centering
\includegraphics[width=\columnwidth]{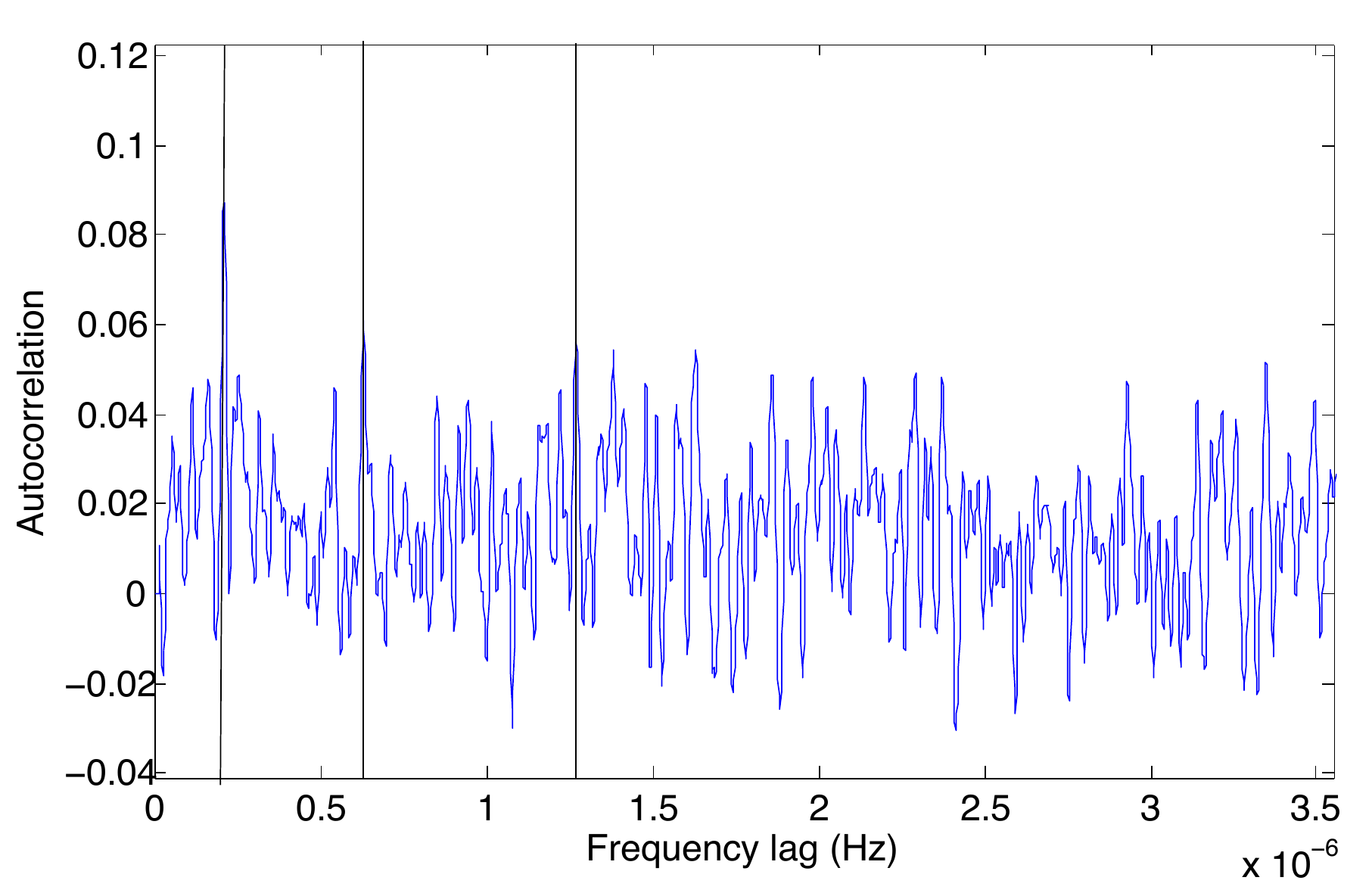}
\caption{Autocorrelation function $\mathrm{A}(\nu)$ of the frequency range 5.5 - 34.7~$\mu$Hz of the power spectrum shown in  Fig.~7.  The first value displayed is bin number 7, i.e., 0.013~$\mu$Hz. The vertical lines are at 210, 630, and 1260~nHz.}
\label{fig:A}
\end{figure}

\begin{figure}
\centering
\includegraphics[width=\columnwidth]{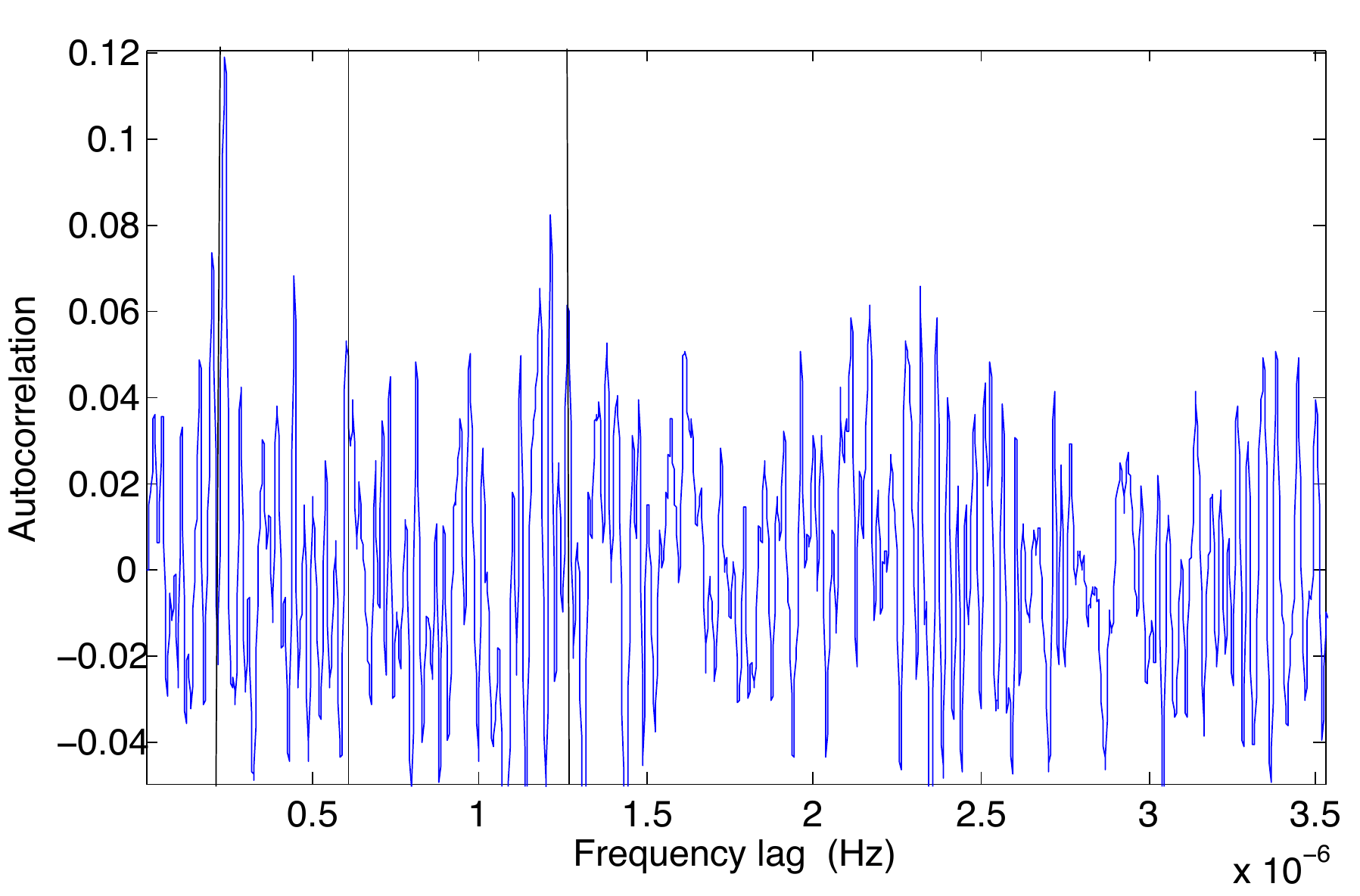}
\caption{Autocorrelation function of the frequency range 5.5 - 13.2~$\mu$Hz of the power spectrum shown in  Fig.~7. }
\label{fig:A_5-13}
\end{figure}

\begin{figure}
\centering
\includegraphics[width=\columnwidth]{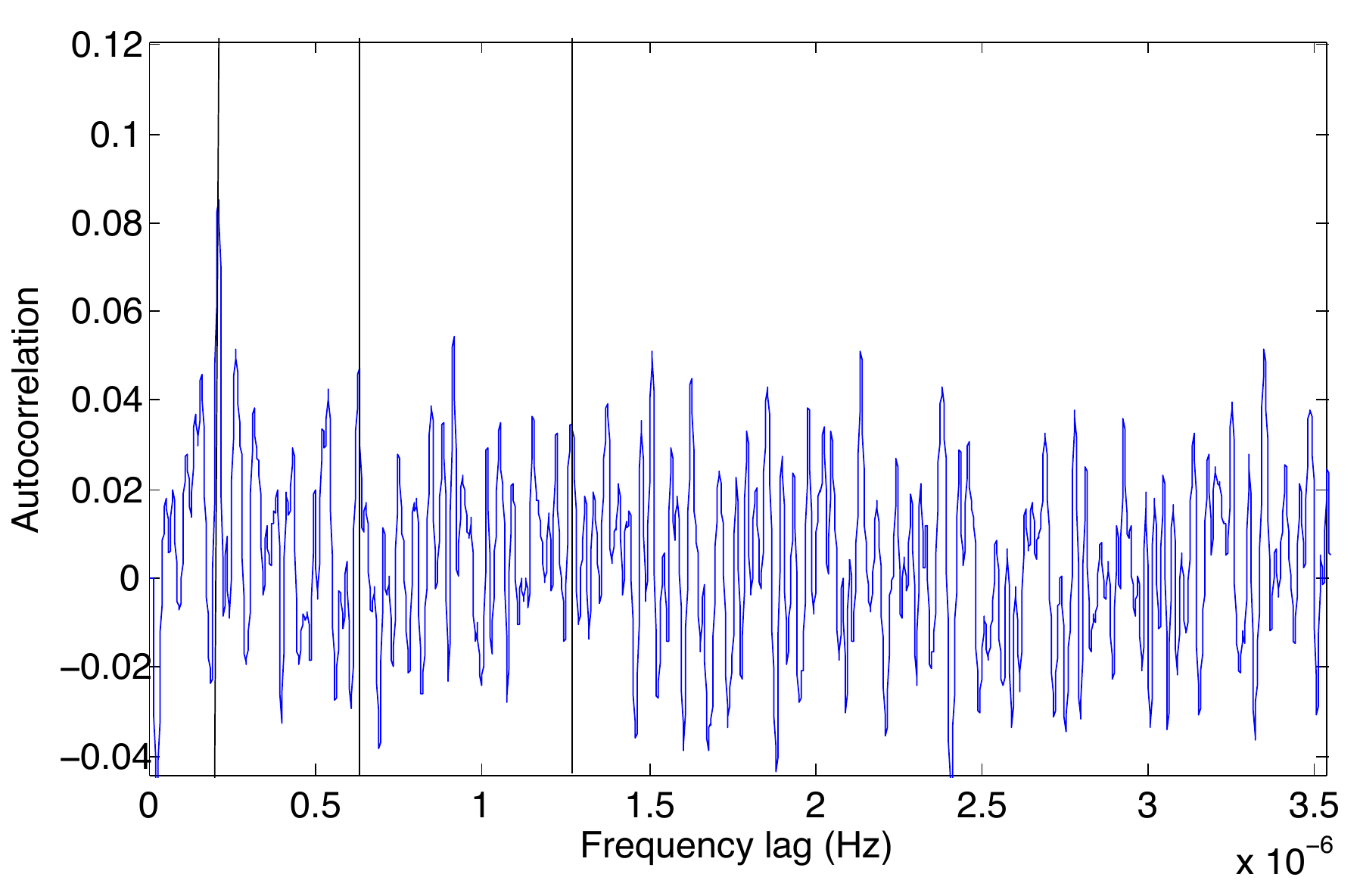}
\caption{Autocorrelation function of the frequency range 13.2 - 34.7~$\mu$Hz of the power spectrum shown in  Fig.~7. }
\label{fig:A_13-34}
\end{figure} 

\begin{figure}
\centering
\includegraphics[width=\columnwidth]{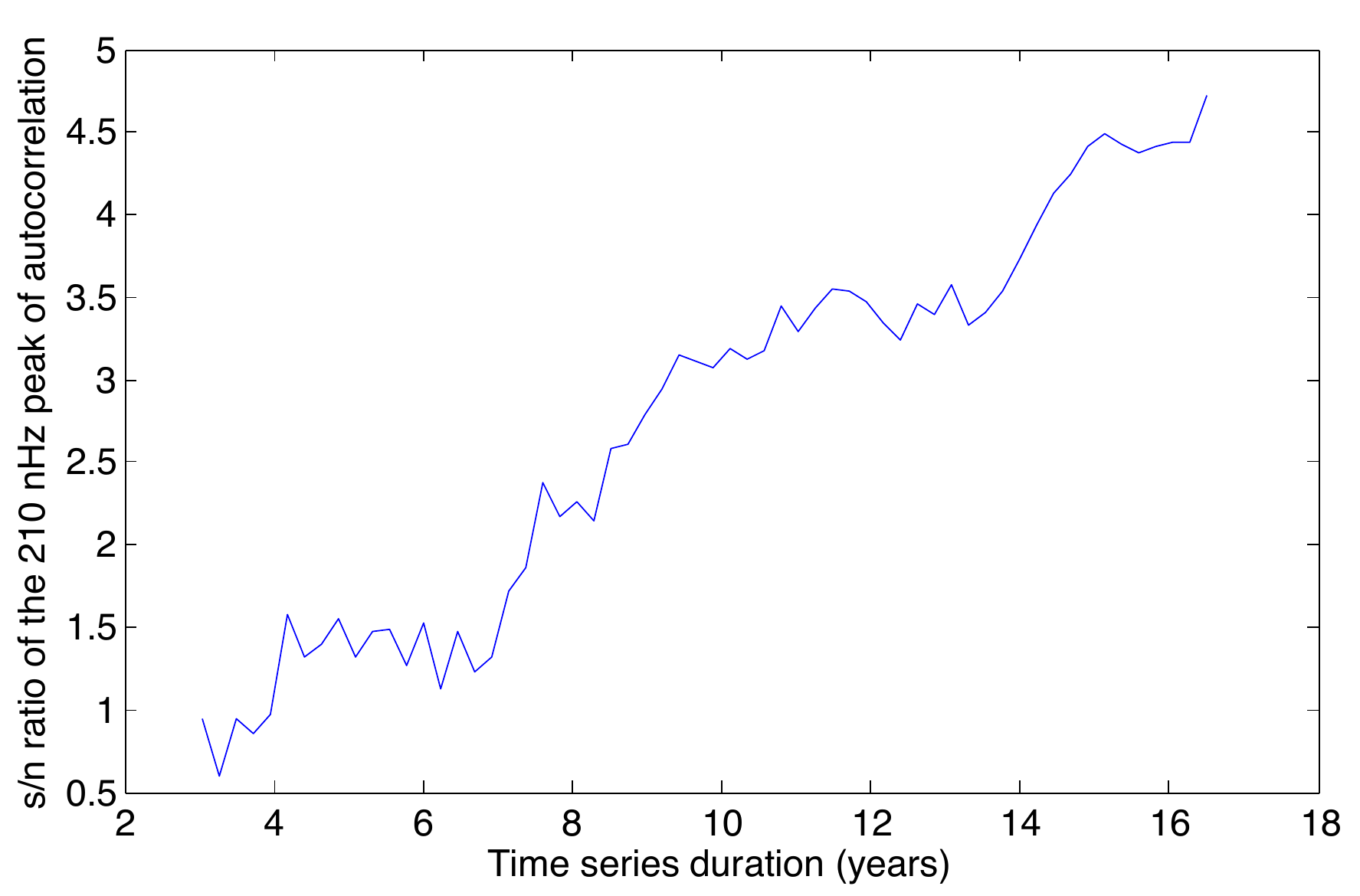}
\caption{S/N ratio of the 210~nHz peak of Fig.~10 as a function of the duration of the time series, from 3 to 16.5 years.}
\label{fig:SN_210peak}
\end{figure}

\begin{figure}
\centering
\includegraphics[width=\columnwidth]{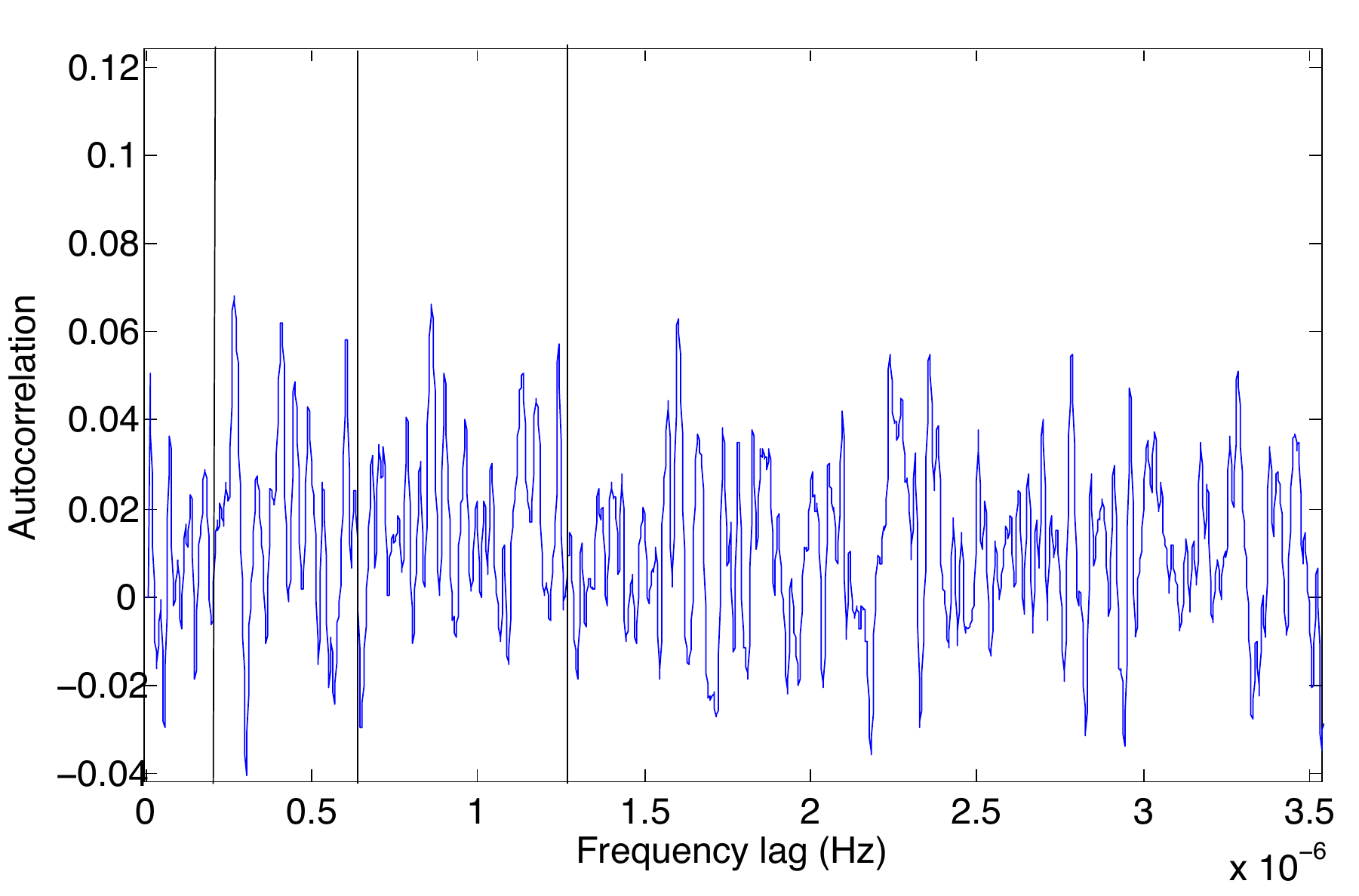}
\caption{Autocorrelation function of the frequency range 5.5 - 34.7~$\mu$Hz of a power spectrum resulting from an artificial data set. No peak is visible at 210~nHz, and the highest peak stands at 3.3~$\sigma$ to be compared to the 4.7~$\sigma$ of the peak obtained from the GOLF data set.}
\label{fig:A_simu}
\end{figure} 

\begin{figure}
\centering
\includegraphics[width=\columnwidth]{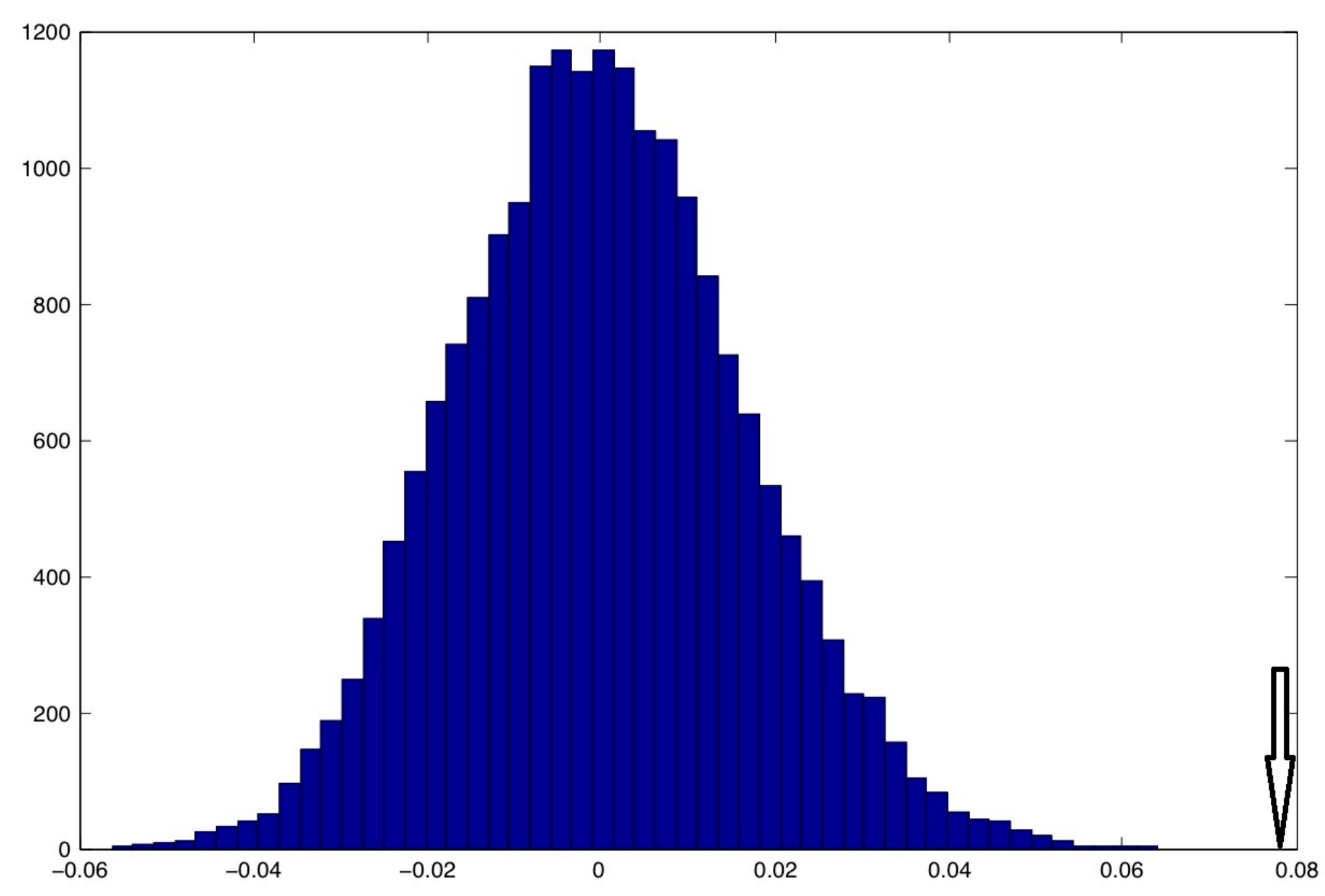}
\caption{Histogram of ten figures similar to Fig.~10 obtained with artificial data, simulated as described in text.}
\label{fig:histo}
\end{figure} 

Before interpreting further Fig. 10, we have to mention that 
special care must be taken here because of the specific process we used for this g-mode detection; our detection agent is the set of low-degree p modes, located inside the Sun and not on Earth or at the Lagrange point. The g-mode signal is then seen from a reference that rotates inside the radiative and convective zones. This will have crucial consequences for the rotation analysis.

When we assume that below the convection zone, the rotation rate depends only on radius $r$ and not 
on the colatitude $\theta$ ($\Omega(r,\theta)=\Omega(r)$ for $r\le r_\mathrm{cz}$), 
then we can define a mean rotation rate sensed by the asymptotic g modes by
\begin{equation}\label{eq:omegag}
\Omega_\mathrm{g} = {\frac{\int_0^{r_{\mathrm{cz}}}{\Omega(r)\left(\frac{N(r)}{r}\right)\mathrm{d}r}}{\int_0^{r_{\mathrm{cz}}}{\left(\frac{N(r)}{r}\right)\mathrm{d}r}}}=\int_0^{r_{\mathrm{cz}}}{K(r)\Omega(r)\mathrm{d}r}
,\end{equation} 
where $r_\mathrm{cz}$ is the base of the convection zone and $N(r)$ is the Brunt-V\"ais\"al\"a frequency.
The sidereal asymptotic splittings are then be given by $\Omega_\mathrm{g}/2$ and $5/6\ \Omega_\mathrm{g}$ for 
$l=1$ and $l=2,$ respectively (Berthomieu \& Provost, 1991).
For low-frequency g modes, below $33\ \mu$Hz with $|n|\ge 20$, the relative difference 
between the mode splittings and these asymptotic values is smaller 
than 3\% (see Fig.~5 of Berthomieu \& Provost 1991).

In this work, the rotational splittings are not seen from an inertial frame. Instead 
we search for them in the signal of the low-degree p modes that 
propagate through the rotating Sun. 
Following Komm et al.~(2003), we can define the mean rotation rate sensed by p modes 
as the solid-body rotation of the Sun given by
\begin{equation}
\Omega_\mathrm{p}={\frac{\int_{r_\mathrm{c}}^{R_\odot}{\rho r^4 \int_0^\pi{\frac{3}{4}\sin^3\theta\ \Omega(r,\theta)}}\ \mathrm{d}r\mathrm{d}\theta}{\int_{r_\mathrm{c}}^{R_\odot}\rho r^4 \mathrm{d}r}}
,\end{equation}
where $\rho$ is the density, $R_\odot$ is the solar radius, and $r_\mathrm{c}\simeq 0.2 R_\odot$ is the boundary below which p-mode splittings become rather insensitive to the rotation rate. From the inversion of MDI splittings, Komm et al.~(2003) reported $\Omega_\mathrm{p}=434.5\pm5.9$~nHz.  In analogy with the synodic-sidereal correction that is necessary when the splittings 
are observed from a terrestrial perspective, the expected asymptotic g-mode splitting components seen in the p-mode signal 
are given for $l=1$ and $l=2$ by

\begin{eqnarray}
s(1,m)&=&m[\frac{\Omega_\mathrm{g}}{2} -\Omega_\mathrm{p}]\label{eq:split1}\\
s(2,m)&=&m[\frac{5\Omega_\mathrm{g}}{6} -\Omega_\mathrm{p}]\label{eq:split2},
\end{eqnarray}

respectively. When we use only the low-degree modes (those used in this g-mode search), no inversion is possible, but a mean rotation rate $\Omega_\mathrm{p}=432.5\pm1$~nHz has been reported by Fossat et al.~(2002). Taking these two estimates into account, we used a mean value of $\Omega_\mathrm{p}=433.5\pm7$~nHz for the rotation of our "instrumental reference" (i.e., the p modes that corotate with the Sun).

Because our instrumental reference is located inside the rotating Sun, Eqs.~(\ref{eq:split1}) and (\ref{eq:split2}) may imply a surprising consequence. If the g-mode kernel rotation  $\Omega_\mathrm{g}$  were on the same order as the p-mode kernel rotation, close to 433~nHz (case of a nearly rigid rotation down to the solar center), then Eq.~(\ref{eq:split1}) would imply a negative value of the splitting $s(1,1)$, which would indeed be equal to half the value of the p-mode rotation. A negative splitting means that the positions of the two tesseral peaks corresponding to $m=\pm 1$ are reversed. It also means that an observer located on the solar surface, with a constant internal rotation down to the solar center, would "see" the g modes rotating in the opposite direction, at half the global rotation rate. As in the case of the Foucault pendulum, this is a non-intuitive consequence of the Coriolis effect, which is itself due to the nearly horizontal motions of the g modes.

Our main peak at about 210~nHz, close  to half of 433~nHz (although significantly different), could correspond to this situation if it is to be seen as a signature of the $s(1,1)$ splitting. This would mean, incidentally, that the $m=0$ g modes are detected by this method. 
Helioseismologists are used to the fact that p modes with an odd value of ($m+l$) are not visible in full-disk observations. This is an observational bias that arises because the visibility is limited to half the total solar surface. In the present case, the "observation" of g modes is made from inside the Sun itself, so that this bias disappears and our measurements are sensitive to all $m$ components.
The same benefit should then exist for the $m=\pm 1$ of $l=2,$ which are also not directly accessible in full-disk velocity p-mode measurements, but are also expected to contribute here in the g-mode detection. 

However, the autocorrelation does not see the difference between a negative and a positive splitting. We are then confronted by an ambiguity between a possible nearly solid rotation producing a splitting $s(1,1)=-210$~nHz and  a much faster g-mode rotation of more than 1200~nHz, which would produce a splitting $s(1,1)=+210$~nHz.

To solve this ambiguity, we need to also detect the signature of the $l=2$ g modes. Their Coriolis coefficient, 5/6, is quite different so that the same 210~nHz splitting of $l=1$ is associated with very different values of the $l=2$ splitting in the two possible solutions. The $l=2$ g-modes' splitting signature is certainly seen with a lower S/N ratio, as  shown in Fig.~10, where no dominant series of peaks can be clearly assigned to them.  However, they are potentially many, and they consist of quintuplets, so that the increased statistical efficiency of the autocorrelation may still offer a chance that some of the non-dominant peaks of Fig.~10 contain their signature.

In the case of slow rotation, the single $s(2,1)$ splitting should be negative, around -60~nHz, while in the second case, its positive value is expected around  630~nHz. No significant peak can be seen in Fig.~10 around 60~nHz, while the second highest peak of this figure indeed stands near 630~nHz. The third highest peak, although not highly significant by itself,  is located near 1260~nHz, and could be the signature of $s(2,2)$. 

On the other hand, the 210~nHz high peak could mathematically also be the signature of any other splitting component such as $s(1,\pm 1)$ or $s(2,1)$. Equations~(\ref{eq:split1}) and (\ref{eq:split2}) must be satisfied, and this will help to reduce the ambiguity.

\begin{figure}
\centering
\includegraphics[width=\columnwidth]{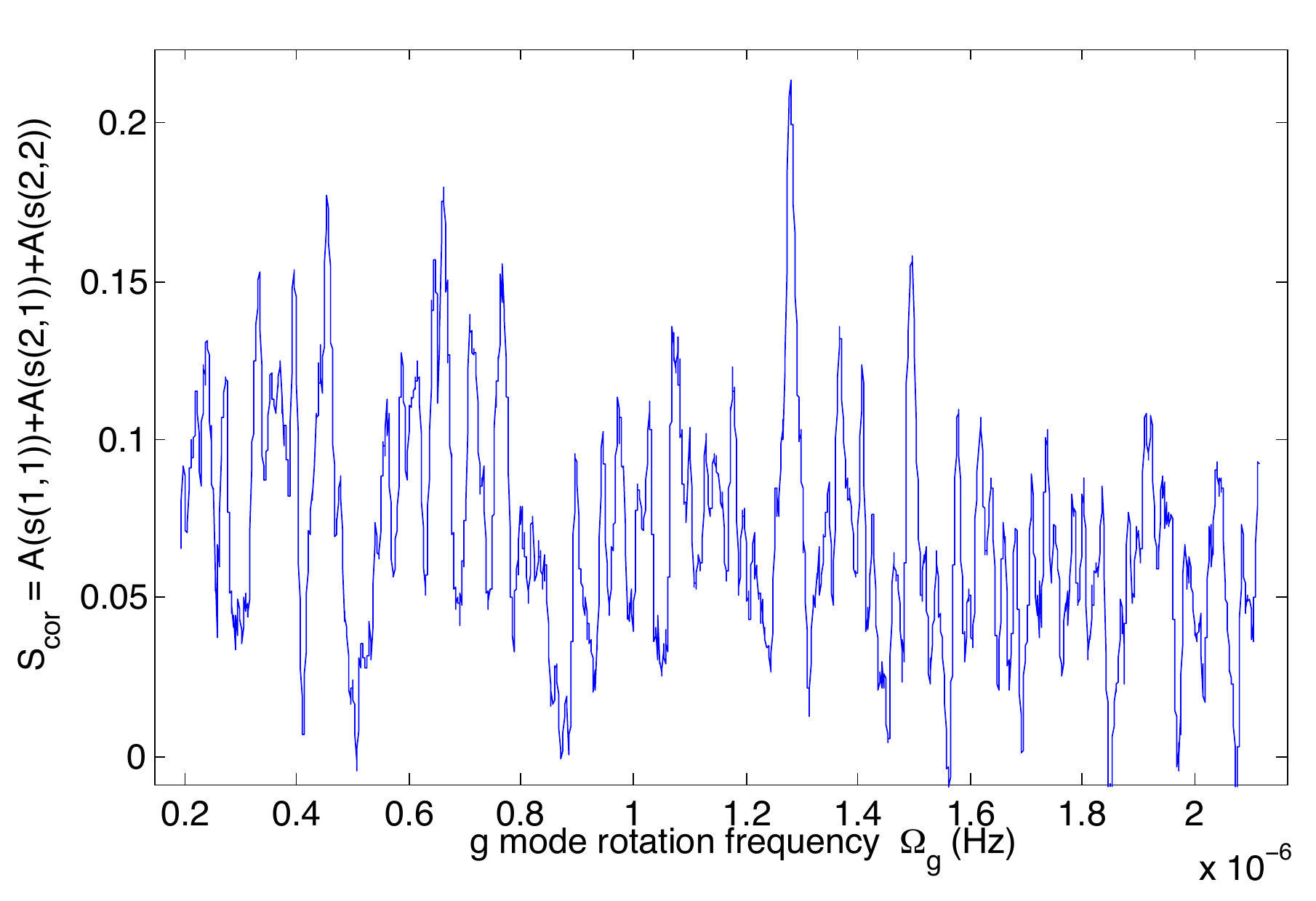}
\caption{Sum $S_\mathrm{cor}$ (Eq.~(\ref{eq:Scor})) of the three possible g-mode splitting signatures in the autocorrelation, as a function of the g-mode kernel rotation $\Omega_\mathrm{g}$. It identifies a fast rotation rate of $1277 \pm 10$~nHz.}
\label{fig:Scor}
\end{figure} 

\begin{figure}
\centering
\includegraphics[width=\columnwidth]{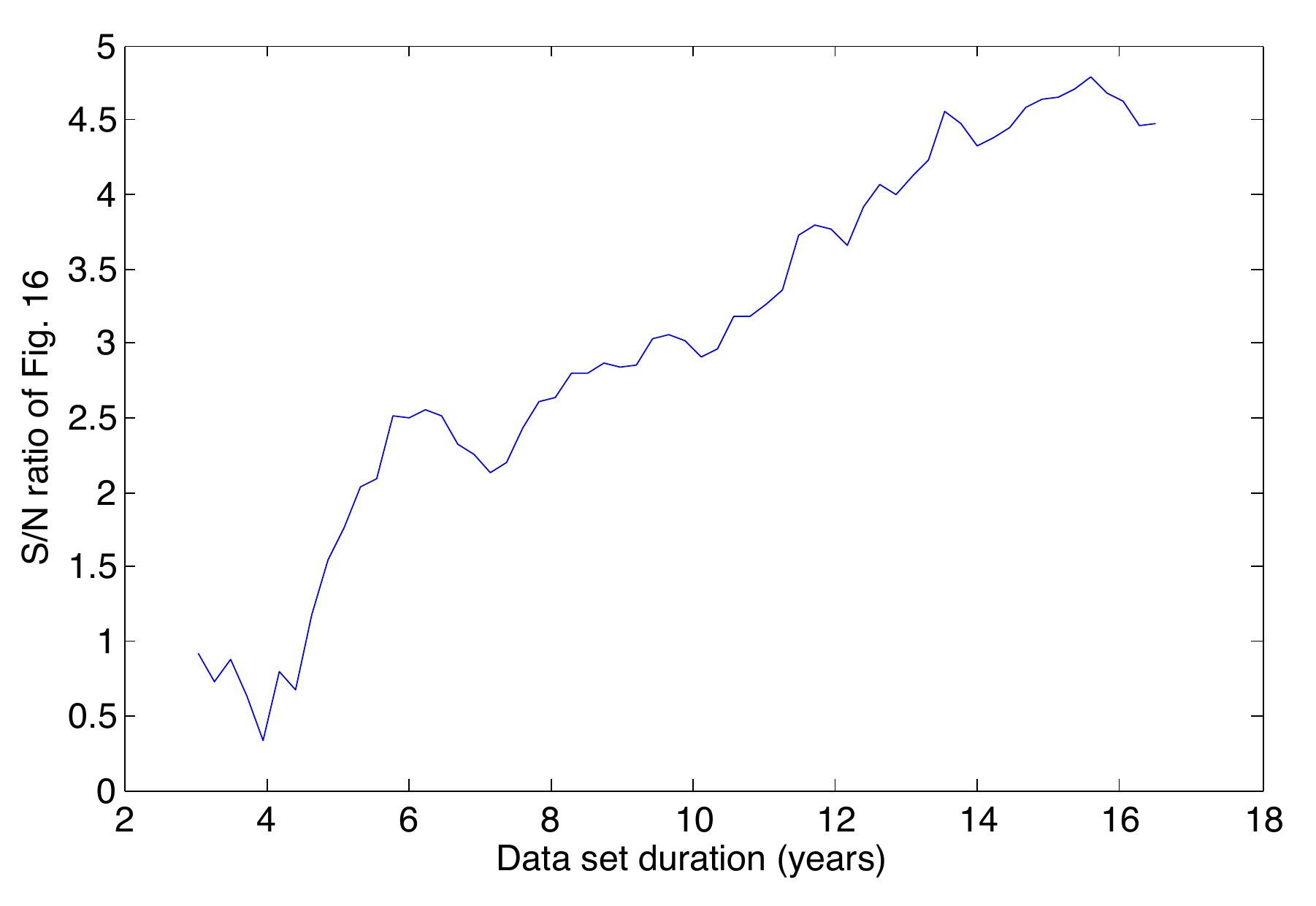}
\caption{S/N ratio of the main peak of Fig.~16 as a function of the duration of the time series analyzed, from 3 to 16.5 years.}
\label{fig:SN_Scor}
\end{figure} 

The simplest way of checking the significance of these various possible signatures of g-mode splittings is to plot the sum of the autocorrelation values at the three abscissae $s(1,1)$, $s(2,1),$ and $s(2,2)$ as a function of  $\Omega_\mathrm{g}$, 

\begin{equation}\label{eq:Scor}
S_\mathrm{cor} = \mathrm{A}(s(1,1))+\mathrm{A}(s(2,1))+\mathrm{A}(s(2,2))
,\end{equation}
with $s(l,m)$ defined by Eqs.~(\ref{eq:split1}) and (\ref{eq:split2}), as abscissae of the ordinate A plotted in Fig.~10.

This is shown in Fig.~16, where the sharp highest peak clearly favors the rapid rotation solution. In this figure,  $\Omega_\mathrm{g}$ is sampled at 1~nHz intervals, and the three abscissae $s(1,1)$, $s(2,1),$ and $s(2,2)$ are rounded to the integer value of the bin number obtained from Eqs.~(\ref{eq:split1}) and (\ref{eq:split2}). An easy Monte Carlo test demonstrates the high significance of this peak: a random selection is made of three abscissae in Fig.~10, and their three ordinates are added. One million such selections can easily be made, and repeating this one hundred times provides on average one occurrence of such a peak at  5.2~$\sigma$ (with respect to a mean value of 0.04 and an r.m.s. scatter of 0.029 in each set of one million selections).  This occurrence always corresponds to contributions of the 210~nHz peak plus two of the  three or four highest other peaks, and only one of these solutions satisfies Eqs.~(\ref{eq:split1}) and (\ref{eq:split2}). Figure~17 shows that the S/N ratio of this g-mode rotation peak at about 1280~nHz increases with increasing data set duration. This increase does not differ much from the increase of the 210~nHz peak alone, an indication that taking the assumed $l=2$ contributions
into account does not only add noise.  This 1280~nHz peak is consistent with the addition of the three highest peaks in Fig.~10, within their respective uncertainties of a few nHz. 

For the statistical significance of the $l=2$ splitting contribution, it is possible to reproduce Fig.~16  by using separately only the contribution of the $l=1$ splitting, or the contributions of the first two splitting values of $l=2$. This is shown in Fig.~18 in blue for $l=1$ and red for $l=2$. The blue curve shows the ambiguity on the rotation rate created by Eq.~(\ref{eq:split1}), which is visible as a symmetry around 866~nHz ($2\times433$), so that two dominant peaks that are equally high support either a slow or a fast rotation. The red curve, corresponding to $l=2$ and Eq.~(\ref{eq:split2}), displays a symmetry around 520~nHz ($433\times 6/5$) and clearly helps selecting one unique solution
that is precisely located on one of the other two peaks. They are slightly rescaled here for better visibility, and Fig.~16 shows the sum of these two curves. The rapid rotation of about $1277\pm10$~nHz could therefore almost be measured by the only contribution of the $l=2$ g modes, but the case becomes very compelling with the two contributions taken together. The error bar of 10~nHz around 1277~nHz takes the three contributions of the three correlation peaks into account. 
We can then conclude that the autocorrelation shown in Fig.~10 indeed contains the significant signatures of g-mode splittings of degrees 1 and 2. This indicates a rapid rotation of the solar
core, and we  can move  to the next step of the analysis, but we return to this important point of the rotation in the general discussion.

\begin{figure}
\centering
\includegraphics[width=\columnwidth]{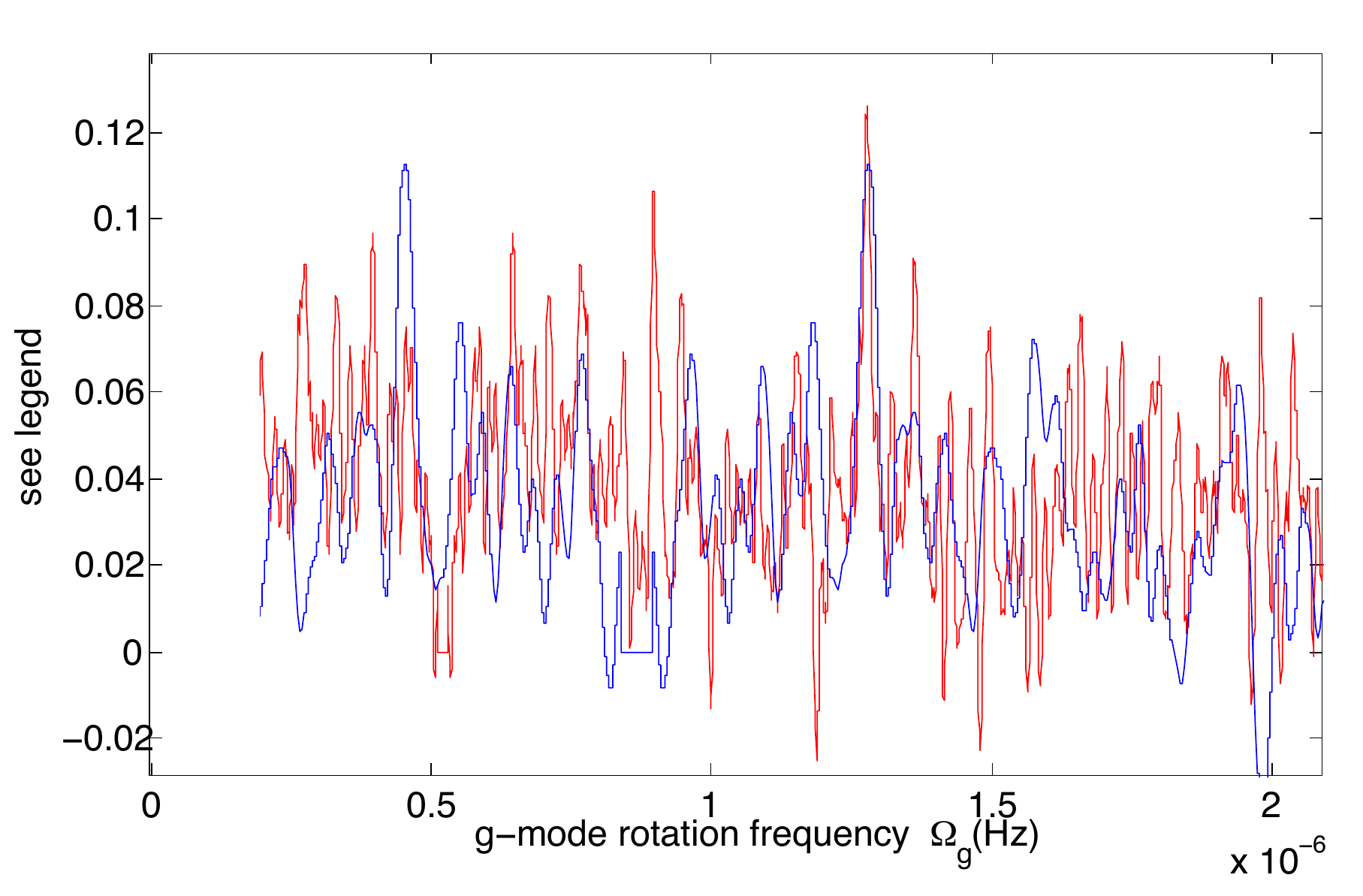}
\caption{Same as Fig.~16, but computed separately for the unique contribution of $l=1$, $\mathrm{A}(s(1,1))$ in blue, or for the contribution of two components of $l=2$, $\mathrm{A}(s(2,1))+ \mathrm{A}(s(2,2))$ in red. This illustrates that the $l=2$ contribution confirms one of the two possible rotation rates detected by $l=1$ with ambiguity. }
\label{fig:Scor_1_2}
\end{figure} 

It is also interesting to compare Fig.~18 with the same calculation made on the same artificial data set, shown in Fig.~19. It is clearly more random.

\begin{figure}
\centering
\includegraphics[width=\columnwidth]{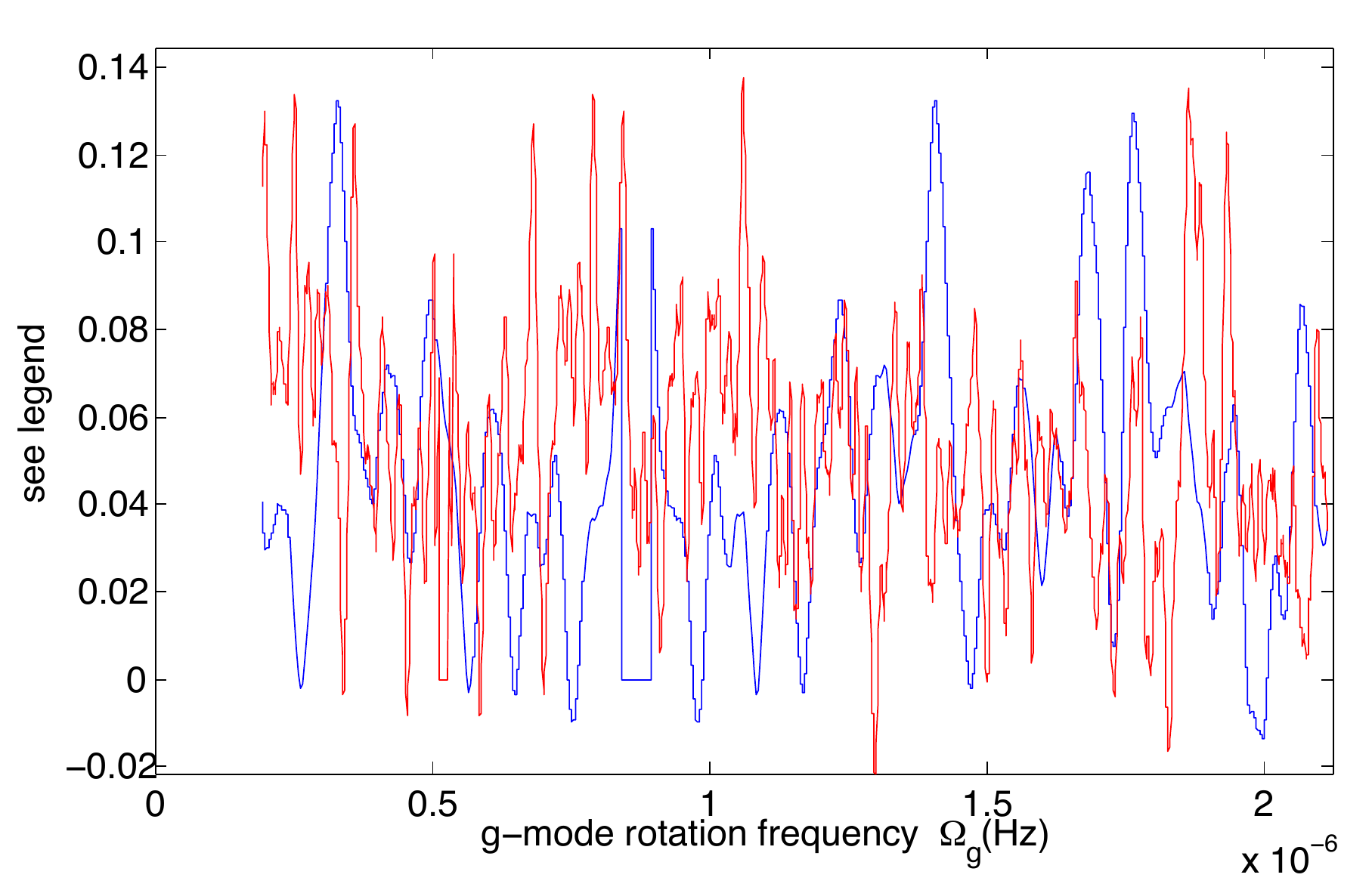}
\caption{Same as Fig.~18 for an artificial data set. }
\label{fig:Scor_1_2_simu}
\end{figure}

\section{Equidistant period spacing. The case of $l=1$}

 For a sufficiently low frequency, the second-order asymptotic period  
is given by
\begin{equation}\label{eq:2ndasymptotic}
P_{n,l}=P_{l}(|n|+l/2+\theta )+O(1/P_{n,l})
,\end{equation}  
where $P_{l}$  is the equidistant period spacing, defined by
\begin{equation}\label{eq:Pl}
P_l=P_0/\sqrt{(l(l+1)}
,\end{equation}
and $\theta$ is a phase factor that tends to a constant when the radial order increases (Berthomieu \& Provost 1991; Provost et al. 2000).
$P_0$ is directly related to the denominator of 
the asymptotic g-mode rotation kernel given by Eq.~(\ref{eq:omegag}), 
\begin{equation}\label{eq:P0}
P_0=\frac{2\pi^2}{\int_0^{r_\mathrm{cz}}{\frac{N(r)}{r}\mathrm{d} r}}.
\end{equation}
P-mode helioseismology has bracketed the value of $P_{0}$  to be around 34 minutes, with an uncertainty smaller than one minute. The two asymptotic values of $P_{l}$ should then be approximately 24 minutes for $l=1$ and 14 minutes for $l=2$.  We separately
address  the case of  $l=1$ , which provides a highly significant result. The case of  $l=2$ is expected to be more difficult.
 
 In this section we assume that all of the g modes participating in our analysis indeed precisely follow this equidistant period spacing asymptotic approximation. We discuss this hypothesis in the general discussion section.

Given the expected equidistant period spacing of about 24 minutes, there must exist about 100 g-mode dipole frequencies in the range explored to produce Fig.~7. We do not know how many of these really contribute to this figure because they are not individually detectable.  We know from Figs.~11 and 12 that they are many,
however. A priori we should explore the broadest possible frequency range to search for the equidistant period spacing. Another difficulty arises, however, which is the density of g modes in the lowest part of this range. Because they are equidistant in period, the consecutive modes approach each other with decreasing frequency. At  6~$\mu$Hz, two consecutive frequencies are separated by about 51~nHz, while their $s(1,1)-s(1,-1)$ splitting is  420~nHz. This means that more than eight consecutive g modes overlap.  This density provides an interesting statistical benefit in the autocorrelation, while it might create confusion in this part of the analysis. 

Provost \& Berthomieu~(1986) and Christensen-Dalsgaard \& Berthomieu~(1991) estimated that the asymptotic approximations could be regarded as valid within about $10^{-3}$. In our frequency range this means a few bins. The 6-bin smoothing we used for the autocorrelation has indeed proved to optimize the visibility of most of the following figures.

The search for equidistant period spacing makes use of the assumed known splitting to produce one model of many multiplets that
cover part of or all the frequency range used to measure the rotation. This model has two free parameters: the equidistant period spacing $P_l$, 
and the position of this equidistant series, which is defined by the value $P_{\mathrm{min},l}$ of the shortest period of the model.

A 2D exploration ($P_l$ in abscissa and  $P_{\mathrm{min},l}$ in ordinate) can be computed
 by adding all of the individual contributions of the spectrum  of $T$ (Fig.~7)
along the abscissae of this model. This can be written as  

\begin{equation}\label{eq:Cl}
C_l(P_l,P_{\mathrm{min},l}) = \sum_{i=0}^{N_l-1}\sum_{m=-l}^l P_\mathrm{S}\left((P_{\mathrm{min},l}+i P_l)^{-1} + m\times s_l\right)
,\end{equation}
where $P_\mathrm{S}(\nu)$ is the value of the power spectrum of $T$ at the frequency $\nu$. The frequencies are approximated to the integer number of bins, which is 
close enough with the 6-bin smoothing used in all the next steps of the analysis. $N_l$ and $s_l=s(l,m)/m$ are respectively the number of modes and the single splitting value used in the model for degree $l$.

For $l=1$, using $s_1=210$~nHz, we started with the broadest possible 
frequency range of 5.5 - 34.7~$\mu$Hz, and then attempted to reduce it 
by starting above 5.5~$\mu$Hz and determining whether this would  increase or 
decrease the  S/N ratio. The best result is obtained with a set of $N_1=76$ modes, 
starting just above 7~$\mu$Hz.

Figure~20 shows $C_1(P_1,P_{\mathrm{min},1}),$ which uses this model, including the three components of 76 assumed $l=1$ g-modes   
triplet $\nu_{n,1}$ and $\nu_{n,1} \pm 210$~nHz.
This shows the resulting summation of these 228 values, displayed for a 120~s range of the equidistant period spacing centered on 24~min, 20~s (abscissa), and a 2000~s time window starting at 30000~s (ordinate axis). The horizontal range is much wider than the estimated uncertainty in the previous knowledge of the period spacing, and the vertical range is wider than this spacing, so that the first two periods of the equidistant series can be seen. The two red spots horizontally define an equidistant period spacing of 1443~s with at worst an uncertainty of 1s, and they are indeed vertically separated by 1443~s. The shortest period of the series is seen at about 30360~s. Figure~21 shows the series of maxima of each line in Fig.~20. This demonstrates that there is no other maximum that can compare to the red spots in the broad range of the 2D figure that covers more than all the possibilities (4.7~$\sigma$ against 2.1~$\sigma$ for the second highest value). It also reveals, quantitatively more precisely than the color view, the first two g-mode periods of the series, at $30360\pm 15$~s and $31803\pm15$~s, respectively, which are indeed separated by the spacing of 1443~s. This is obtained by construction because, as the above formula shows, replacing $P$ by $P+P_1$ leads to a replacement of $P_\mathrm{S}((P + i P_1)^{-1})$ by $P_\mathrm{S}((P + (i+1) P_1)^{-1})$ and thus to a shift of the sum by one term only.   When we assume that the asymptotic regime has been reached in this frequency range and by comparing these values with the frequency table of Mathur et al.~(2007), these peaks, corresponding to frequencies of  $32.94\pm 0.015\ \mu$Hz and  $31.44\pm 0.015\ \mu$Hz, can tentatively be identified with the signatures of dipolar g modes of radial orders $n=-20$ and $n=-21,$ that is, $P_{\mathrm{min},1}=P_{-20,1}$ in Eq.~(\ref{eq:Cl}). They might differ slightly from the real frequencies of these two modes if the strict asymptotic regime starts beyond the 20th radial harmonics. We return to this point in the general discussion.

\begin{figure}
\centering
\includegraphics[width=\columnwidth]{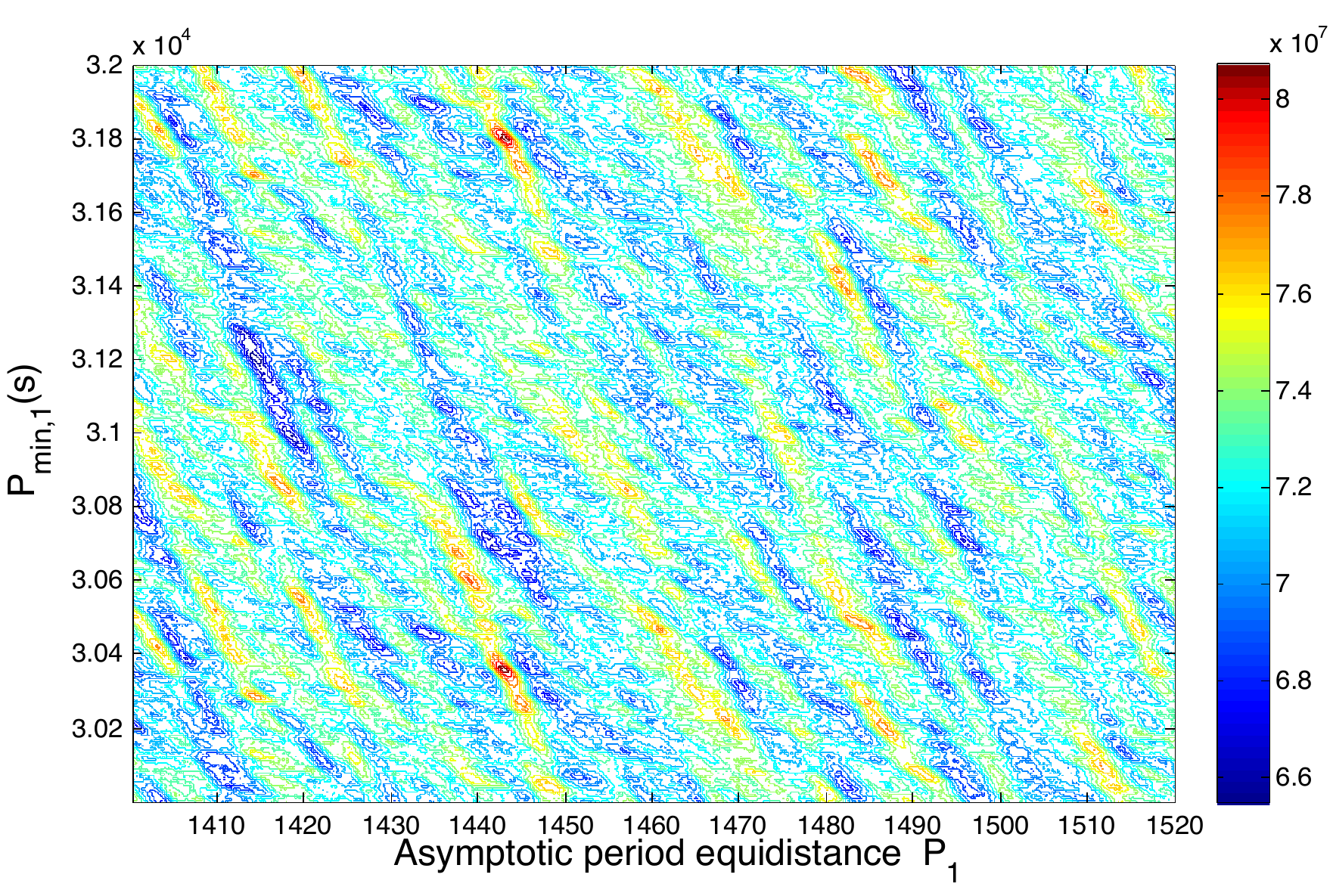}
\caption{Sum $C_1(P_1,P_{\mathrm{min},1})$ (Eq.~(\ref{eq:Cl})) of the 228 values of  a model based on 76 g-mode dipole triplets with the splitting value $s_1=210$~nHz, applied to the power spectrum of Fig.~7, smoothed over 6 bins. The two free parameters of the model are the equidistant period spacing $P_1$, defining the abscissa, and the period $P_{\mathrm{min},1}$ of the first selected mode, defining the ordinate axis. The first (lowest) sharp peak is located at $30360\pm 15$~s (FWHM) and covers the radial order range from -20 to -95, the second (upper part of the figure) covers the radial order range from -21 to -96. The equidistant period spacing is 1443~s, or 24~min~03~s,  within less than 1~s.}
\label{fig:C1}
\end{figure} 
 
\begin{figure}
\centering
\includegraphics[width=\columnwidth]{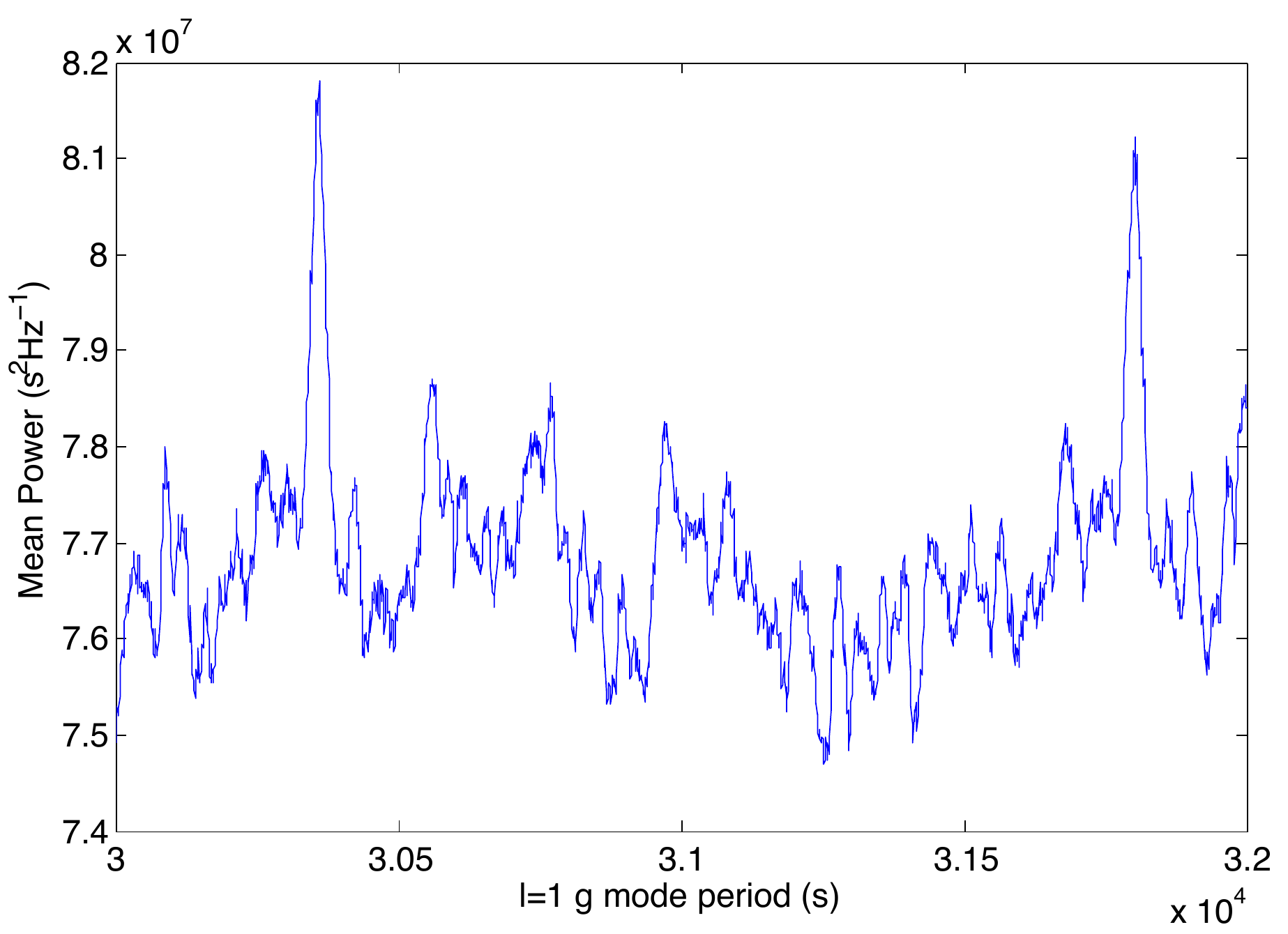}
\caption{Each value is the maximum of a row of Fig.~20, so that the abscissa is the ordinate of Fig.~20. The interval between these peaks is the $l=1$ equidistant period spacing of 1443~s. The power scale is the mean power per individual g-mode peak.}
\label{fig:max_row_C1}
\end{figure} 

At this step it is useful to make a comparison with the same analysis made on an artificial data set. The equivalent of Fig.~20 with the artificial time series we used earlier is shown in Fig.~22, where no specific period equidistance is clearly visible. The equivalent of Fig.~21, shown in Fig.~23, confirms  that it is not the analysis that produces the results. 
We note that Figs.~21 and 23, as well as the next figures of the series (Figs.~24, 25, 27, and 35) are all presented with the same power unit of s$^2$Hz$^{-1}$, as the original spectrum of Fig.~7, so that the amplitudes of the displayed peaks are directly comparable. 

Given the very high S/N ratio of Fig.~21, the idea  of  independently
carrying out the two steps of this analysis (rotation on one hand, equidistance on the other hand)  can be attempted by assuming that the rotational splitting is unknown and by assuming a contribution of the central component of the split triplet (the signature of the simple splitting $s(1,1)$ would not be visible in the autocorrelation otherwise). Then, the search for the equidistant period spacing can be made by the same analysis by reducing the model to a simple series of equidistant central periods, instead of the triplets. We have to expect a loss of sensitivity of about  $\sqrt3$, but it can still be successful because these peaks are at 4.7$\sigma$. The result is shown in Fig.~24. The S/N is indeed significantly reduced, but the series of equidistant periods starting at 30360~s is still clearly identified from the central g-mode frequencies alone. 
After the 210~nHz peak in the autocorrelation, this is a second and independent strong indication of a signature of asymptotic g modes of degree 1 in the power spectrum of Fig.~7. 

We can conclude from this exercise that the central components of the g mode indeed clearly contribute to the previous figures, and we can try to proceed one step further and directly detect their mean contribution in the power spectrum of $T(t)$ itself. We cannot see the triplets individually, but since we know their 76 periods, we know 76 individual frequencies and can compute the mean of 76 sections of the Fig. 7 power spectrum, each one centered on one of these 76 mode frequencies and broader than the total splitting of $s(1,1)-s(1,-1)$. The statistical benefit for each one of the three tesseral components is comparable to what  was shown in Fig.~24, where we were searching for the best solution for efficiently adding 76 individual central components. Now we have this best solution (the positions of the 76 periods, equidistant by 1443~s), and we test whether their addition in the frequency domain will also reveal the two tesseral splitting components. The following formula was used to derive Fig.~25, over the limited range of 1.2~$\mu$Hz: 

\begin{equation}\label{eq:M}
\mathrm{M}(\nu)=\sum_{n=-20}^{-95} {P_\mathrm{S}(\nu_{n,1}-\nu)}
,\end{equation}
where $\nu$ is the frequency and $\nu_{n,1}$ is the inverse of $P_{n,1}=P_{-20,1}+(|n|-20)\times P_1$  (rounded to the integer value of the bin number). 

Figure~25 exhibits the two split tesseral components with an S/N that can be compared to the S/N of Fig.~21, and with a splitting value that is indeed consistent with the 210~nHz value seen in the autocorrelation, and which can be here estimated at $209.5\pm 2$~nHz. Figure~25 also shows that the central component does not only contribute, but contributes with an amplitude about 1.5 times larger than the other two components. This will have to be understood, and it explains, or justifies, the fact that the double splitting $s(1,1)-s(1,-1)$ is not detected in the autocorrelation: with these relative amplitudes, its signature must be three times smaller, so that it can easily disappear in the background noise. 

It is also interesting to note that the mean heights of these peaks can be translated into the modulation amplitude of the sound wave travel time through the solar diameter. The product of height $\times$ line-width is the square of this amplitude. With a mean height of $10^{7}$~s$^2$Hz$^{-1}$ and a line width of about 10~nHz, this results in a  modulation amplitude of about  $\sqrt{10^{-1}}\simeq 0.3$~s, or as a fraction of the travel time of 4 hours, a relative amplitude modulation of $2\times 10^{-5}$.

\begin{figure}
\includegraphics[width=\columnwidth]{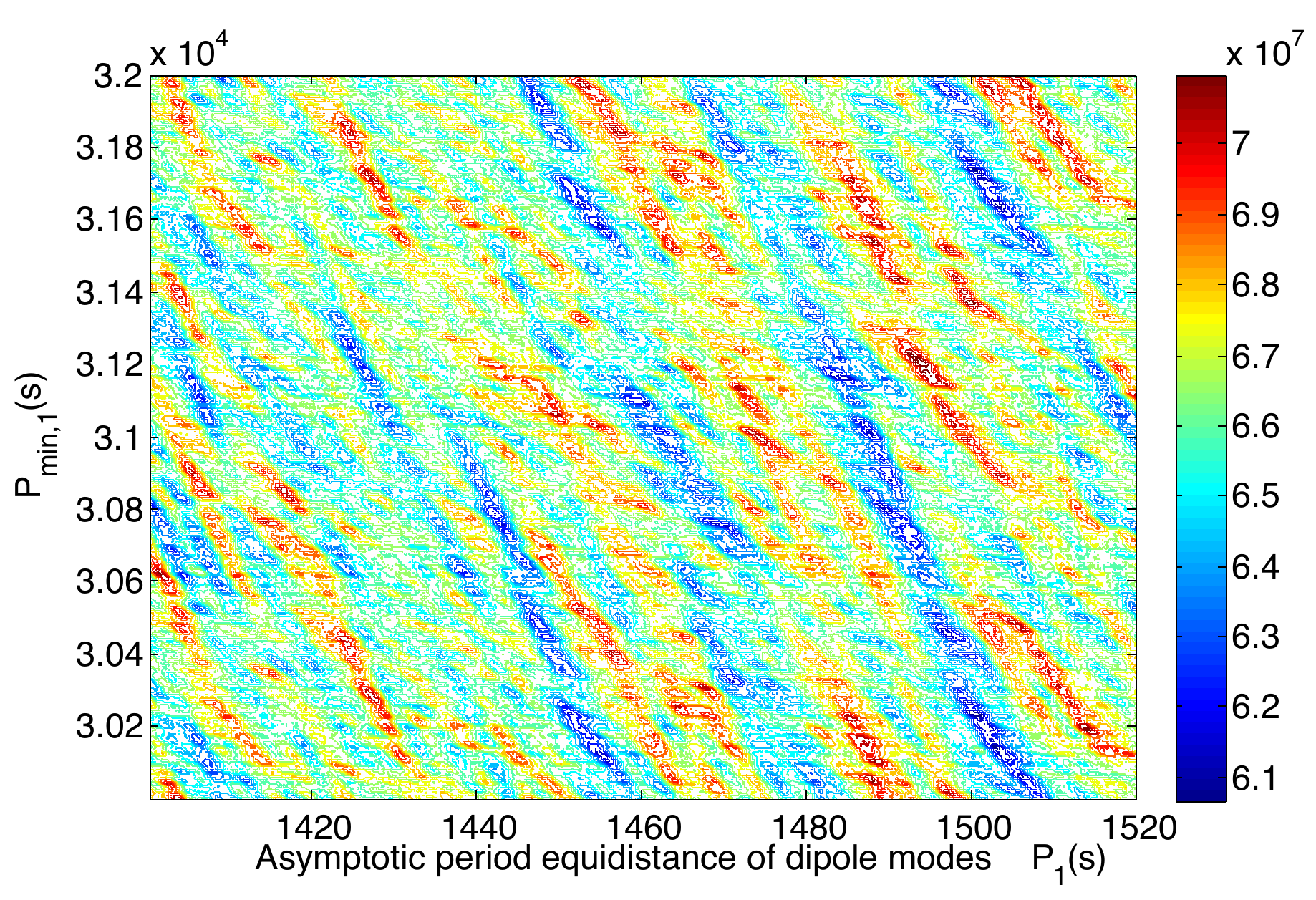}
\caption{As in Fig.~20, the sum of the 228 values of  a model based on 76 g-mode dipole triplets and applied on an artificial data set, with an assumed splitting value of 267~nHz that corresponds to the highest correlation peak in Fig.~14. No significant equidistant period is shown, as demonstrated quantitatively by the next figure.}
\label{fig:C1_simu}
\end{figure} 

\begin{figure}
\includegraphics[width=\columnwidth]{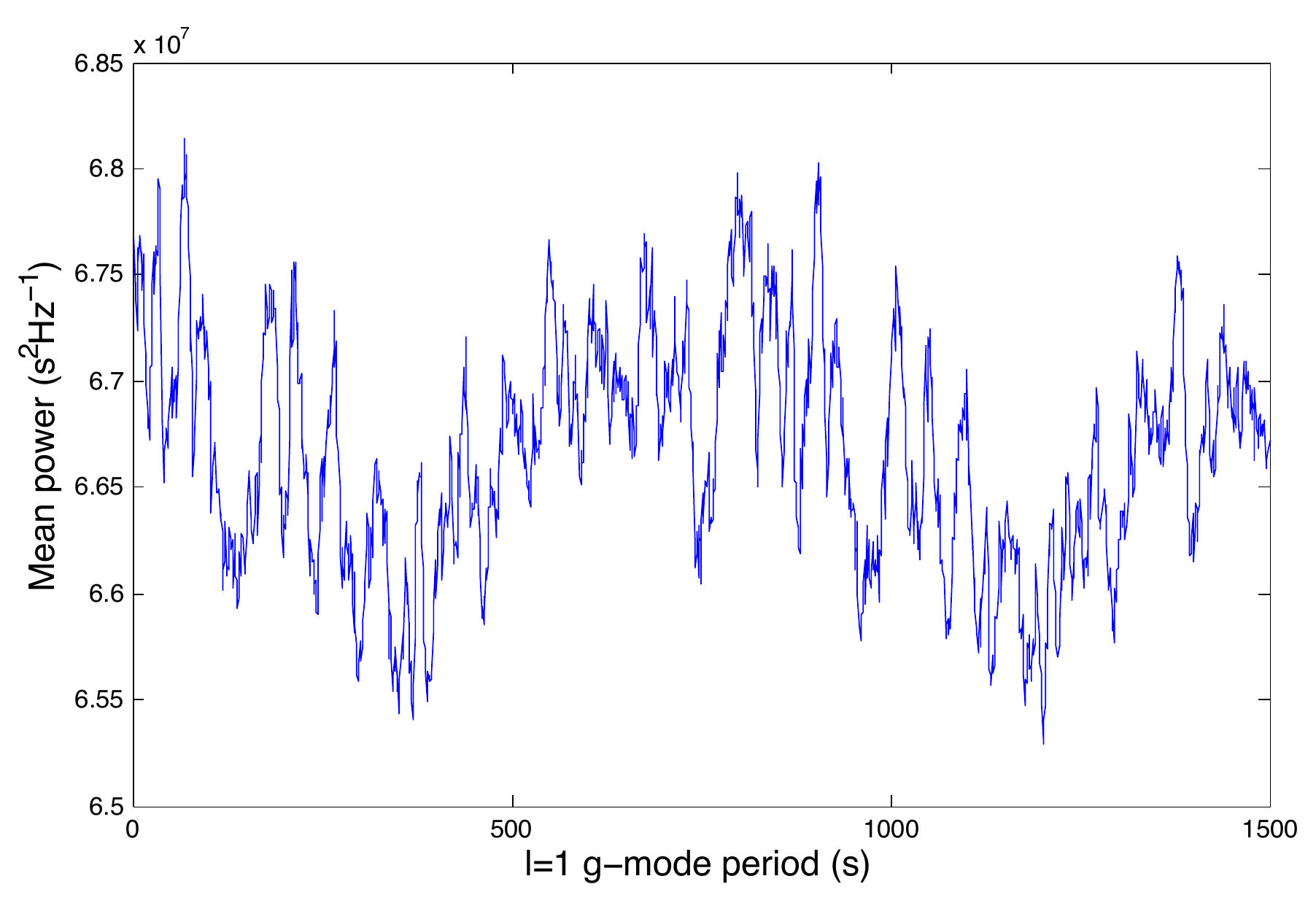}
\caption{Each value of this figure is the maximum of a row of Fig.~22, so that its abscissa is the ordinate of Fig.~22.  This figure relative to an artificial data set must be compared to Fig.~21 obtained with the real GOLF data.}
\label{fig:max_row_C1_simu}
\end{figure}

\begin{figure}
\centering
\includegraphics[width=\columnwidth]{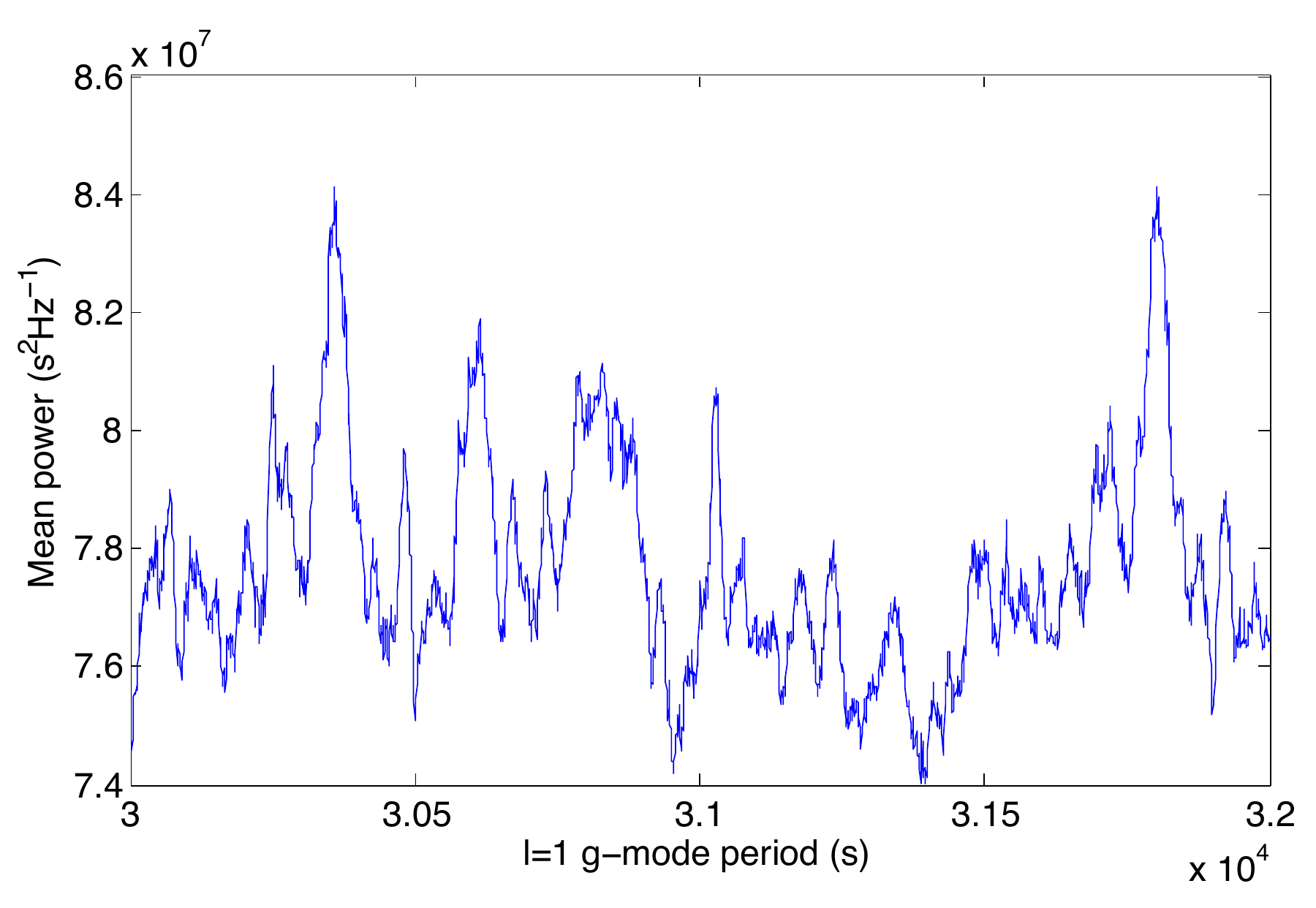}
\caption{   Sum of the 76 values based on the same model based on 76 g modes; this model is simplified to the 76  central g-mode frequencies, thus ignoring knowledge of their splitting.}
\label{fig:sum_central}
\end{figure}

\begin{figure}
\centering
\includegraphics[width=\columnwidth]{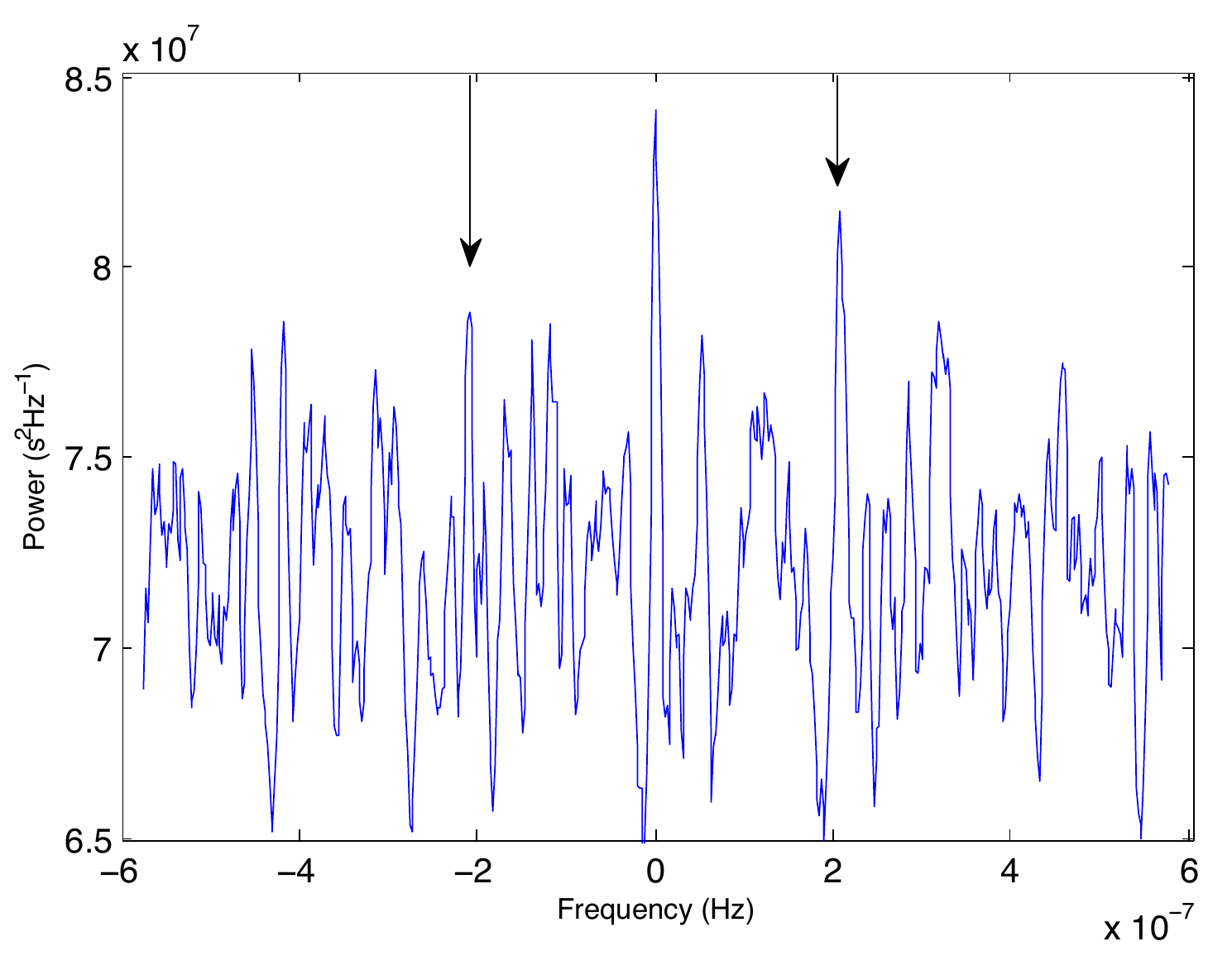}
\caption{Mean profile $\mathrm{M}(\nu)$ of 76  equidistant (in period) dipole modes whose central frequencies are defined by Fig.~20, shown for a $1.2\ \mu$Hz bandwidth (Eq.~(\ref{eq:M})). The triplet splitting of this profile, even when contaminated by the background noise and many overlapping neighbors, is nevertheless directly measured on the modes themselves, at $209.5\pm 2$~nHz.}
\label{fig:M}
\end{figure}

\section{Equidistant period spacing. The case of $l=2$ }

The same general analysis can be applied to the search for the quadrupole modes. This is of course significantly more difficult because of  the smaller detected amplitudes. In the autocorrelation, the main peak of the $l=1$ splitting signature is the sum of (at most) 200 products of individual peaks corresponding to $m=0$ and $m=\pm 1$. In the same frequency range, there are about 170 frequencies of quadrupolar modes, composed of quintuplets. The first signature of their splitting in the autocorrelation is then the sum, at most, of $4\times 170 = 680$ similar products of individual peaks corresponding to neighbor tesseral orders. Despite this encouraging ratio of 680 versus 200, the mean signature in the autocorrelation is about twice smaller (relative heights of the peaks at 630  and 210~nHz), which indicates a mean amplitude, individual mode by individual mode, smaller by at least a factor 4. An important additional difficulty is the very dense overlap of these modes, especially in the lowest frequency range.  They are closer to each other than the dipole modes by a factor $\sqrt3$, while their total splitting width is broader by a large factor: $2520/420 = 6$. This makes it difficult to explore the very lowest part of the frequency range (where each quintuplet is contaminated by the contribution of more than 80 neighbors) and thus reduces  the statistical benefit.

On the other hand, we have in hand the knowledge of the splitting and even of the equidistance period spacing, which is firmly constrained by the asymptotic equation Eq.~(\ref{eq:Pl}),
which implies a value close to $P_2 = P_1/\sqrt3\simeq 833$~s.
We can then attempt the search for equidistant period spacing in a limited range near the expected value of 833~s.

\begin{figure}
\centering
\includegraphics[width=\columnwidth]{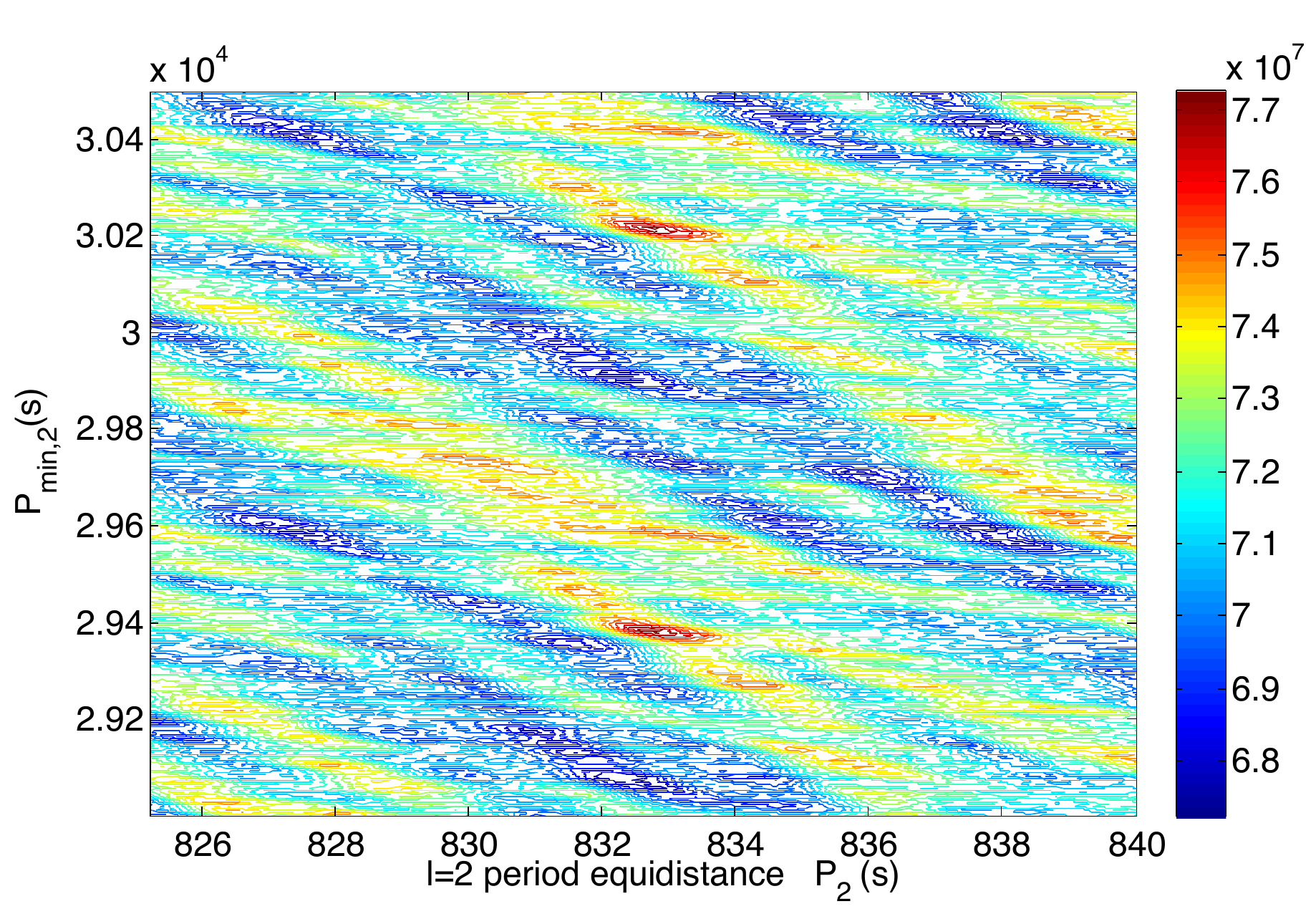}
\caption{Sum $C_2(P_2,P_{\mathrm{min},2})$ (Eq.~(\ref{eq:Cl})) of the 560 components of 112 $l=2$ quintuplets, with a single splitting value of $s_2=630$~nHz, and the two free parameters are the period spacing $P_2$ in the abscissa, in the range 825 - 840~s, and the first central period of the series $P_{\mathrm{min},2}$ in the ordinate, as in Fig.~20. The equidistance is seen at 832.8~s. }
\label{fig:C2}
\end{figure} 

In the case of Fig.~20 for $l=1$, we used a g-mode model of 76 triplets covering the frequency range from 7 to 33~$\mu$Hz. This does not work with $l=2$ because of the large overlap near 7~$\mu$Hz. In an attempt to find a compromise between the need for a good S/N and the excess of confusion created by the overlap of the
modes, we tried here the frequency range starting at 8~$\mu$Hz. Figure~26 is then similar to Fig.~20. It shows $C_2(P_2,P_{\mathrm{min},2})$ (Eq.~(\ref{eq:Cl}) for $l=2$) for a model of $N_2=112$ $l=2$ g-modes quintuplets and a single splitting value of $s_2=630$~nHz. This quantity is displayed over a range of possible equidistance values $P_2$ between 825 and 840~s, and a range of 1500~s for the first value $P_{\mathrm{min},2}$ of the 112 periods, starting  at 29000~s. We can indeed see a possible signature close to the expected equidistance of 833~s, with the visibility of the first two periods separated by about 833~s.


Figure~27 is then a vertical cut of Fig.~26 at 832.8~s, extended to the first three periods. It confirms the equidistant period spacing of 832.8~s with an S/N ratio of 3.65 and also shows the beginning of the decrease in efficiency of the method when pushing it too far toward very long periods.  We show more reasons to accept this $l=2$ detection in the
general discussion.

When we again compare this with the same theoretical frequency table of Mathur et al.~(2007), the first mode of our model can tentatively be identified with the radial order $n = -33$ (i.e., $P_{\mathrm{min},2}=P_{-33,2}$ in Eq.~(\ref{eq:Cl})) at its period of $29380\pm 30$~s, and thus a frequency of $34.04\pm 0.03\ \mu$Hz. We return to the error bars in the general discussion. 

Once again, the same analysis can be made on the artificial data set, and Fig.~28 confirms that it does not show any signature of period equidistance.

\begin{figure}
\centering
\includegraphics[width=\columnwidth]{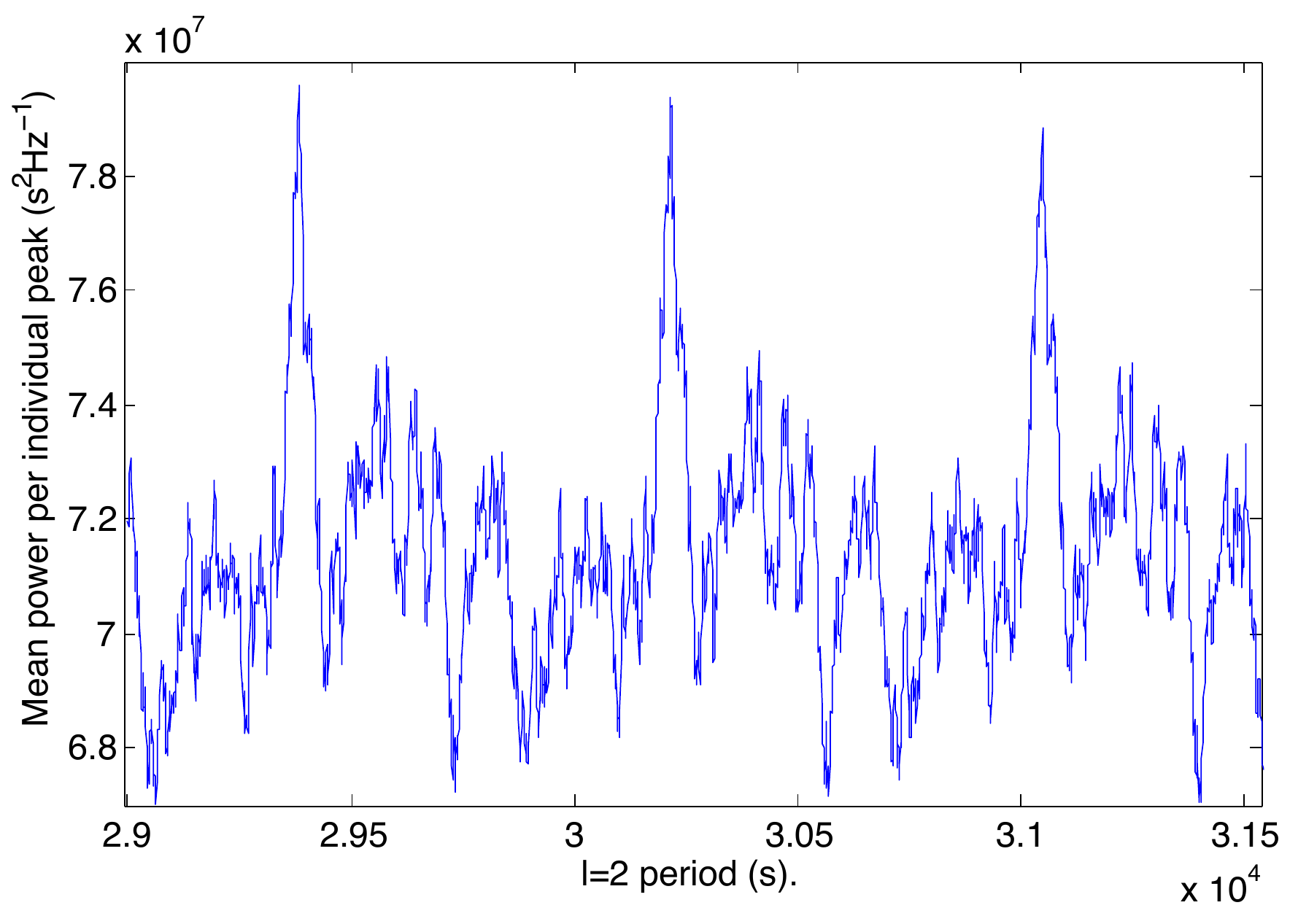}
\caption{Vertical cut of Fig.~26 at 832.8~s, extended to a third period. It shows a good S/N ratio and clearly locates the first period of the series at $29380\pm 30$~s. This peak is the mean of the 560 components of the 112 $l=2$ g modes in the range of radial orders from -33 to -144. The other two peaks cover the ranges (-34,-145) and (-35,-146).}
\label{fig:C2_cut}
\end{figure}

\begin{figure}
\centering
\includegraphics[width=\columnwidth]{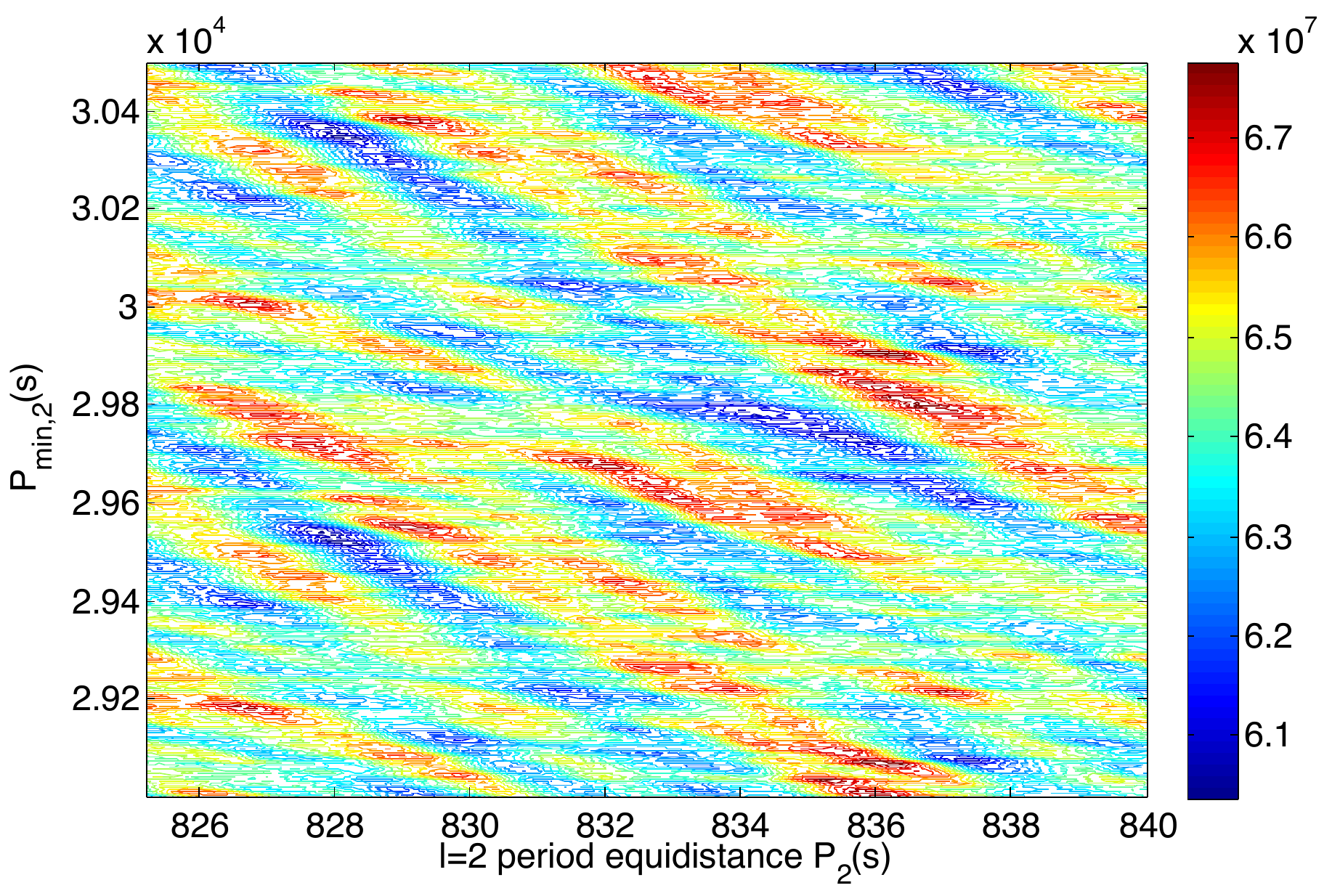}
\caption{Same as Fig.~26 for the artificial data set. This cut is naturally random as no $l=2$ splitting was identified with the simulated data set. Consequently,  no equidistant period spacing can be identified.}
\label{fig:C2_simu}
\end{figure}

\section{Discussion}

The reliability of important results such as the solar g-mode detection and the solar core rotation measurement requires a strong statistical and physical case.  The statistical significance of all results presented so far is comfortable, and we have also shown that the correlation peak at 210~nHz can be obtained from independent data sets  in two non-overlapping frequency ranges. We discuss the solar core rotation first, and then show that independent results can also be obtained for the period spacing equidistances.

\subsection{Rotation of the solar core}

 An interesting additional exercise can be carried out here to reinforce, were it needed, the reliability of this measurement. It consists of considering the problem in the opposite direction. We assume that the autocorrelation does contain the signatures of the three g-mode splittings $s(1,1)$, $s(2,1),$ and $s(2,2)$ and that we do not know their frequencies, they just have to follow, all together, Eqs.~(\ref{eq:split1}) and (\ref{eq:split2}). We do not know  the value of $\Omega_\mathrm{g}$, and this is the most interesting part of this exercise. Then we take the p-mode rotation  $\Omega_\mathrm{p}$ as the unknown parameter of this exercise. We scan a broad range of  $\Omega_\mathrm{p}$ possibilities, and for each possibility we scan the values of $\Omega_\mathrm{g}$ that cover the range of Fig.~16, and we select the highest value of the sum Eq.~(\ref{eq:Scor}), and we finally plot this highest sum as a function of $\Omega_\mathrm{p}$. We still do not need to know $\Omega_\mathrm{g}$. The goal is to determine the best value of $\Omega_\mathrm{p}$ that is consistent with Eqs.~(\ref{eq:split1}) and (\ref{eq:split2}). In other words, we use the only assumption of the  existence of g modes as a tool for measuring the rotation rate sensed by the p modes.  Figure~29 shows that the g modes indeed confirm the 433~nHz rotation rate of the p modes, just a little less accurately than their direct surface measurements. 
We already know (from Figs.~16 and 18) that this $\Omega_\mathrm{p} = 433$~nHz corresponds to $\Omega_\mathrm{g} = 1277$~nHz. We show here that taking Eqs.~(\ref{eq:split1}) and (\ref{eq:split2}) together,
applied to the autocorrelation of Fig.~10, without any a priori knowledge of $\Omega_p$ or $\Omega_\mathrm{g}$, provides an optimal solution with the pair $\Omega_\mathrm{p} = 433$~nHz and $\Omega_\mathrm{g} = 1277$~nHz. As 433~nHz is already well known to be the mean global p-mode rotation, this confirms the strong confidence in the measurement of the g-mode mean rotation $\Omega_\mathrm{g}$ = 1277~nHz.
The uncertainty of this rotation rate is due to an uncertainty in both the measured g-mode splittings and the 433~nHz p-mode rotation. They are both very small, and the total does not exceed $\pm 10$~nHz, that is, the uncertainty is smaller than 1 percent.
 
 The mean rotation rate sensed by the asymptotic g-modes 
$\Omega_\mathrm{g}=1277\pm 10$~nHz  is a weighted average below the convection zone ($r\le r_\mathrm{cz}$).
The weighting function is given by the kernel $K(r)$ (Fig.~30). However, the p-mode splittings have 
already given a good estimate of the mean rotation rate in the radiative interior  
$r_\mathrm{c}\le r\le r_\mathrm{cz}$.  In this  zone, all results published so far are compatible with a 
radiative zone that rotates rigidly  at a mean rate of $433\pm 10$~nHz. 
The area of the kernel $K(r)$ above $r_\mathrm{c}=0.2 R_\odot$ is $30\pm 1\%,$ which leads to an estimate of 
the  mean value of the rotation rate  below $r_\mathrm{c}$ of 
$1644\pm 23$~nHz  (or one-week period), that is, a mean rotation of the solar core $3.8\pm0.1$ times faster than the rotation of the mean radiative zone.

A rapid rotation of the solar core has been suggested by  Garc\'ia et al.~(2007), who also gave an approximate estimate of the $P_1$ value with a precision much lower than that of our analysis: rotation between three and five  times faster than that of the radiative zone, and  $P_1$ between 22 and 26 min (obtained in a higher frequency range not yet very close to the asymptotic period equidistance). Our more accurate results do not contradict the $P_1$ estimate, but they cannot be reconciled with the more precise values of the rotation given later by Garc\'ia et al.~(2011), which were based on the possible identification of five dipolar g modes: the claimed high splittings of these modes, although suffering large scatter,  would imply a g-mode rotation of more than 1800~nHz, to be compared to our result of $1277\pm 10$~nHz.

\begin{figure}
\centering
\includegraphics[width=\columnwidth]{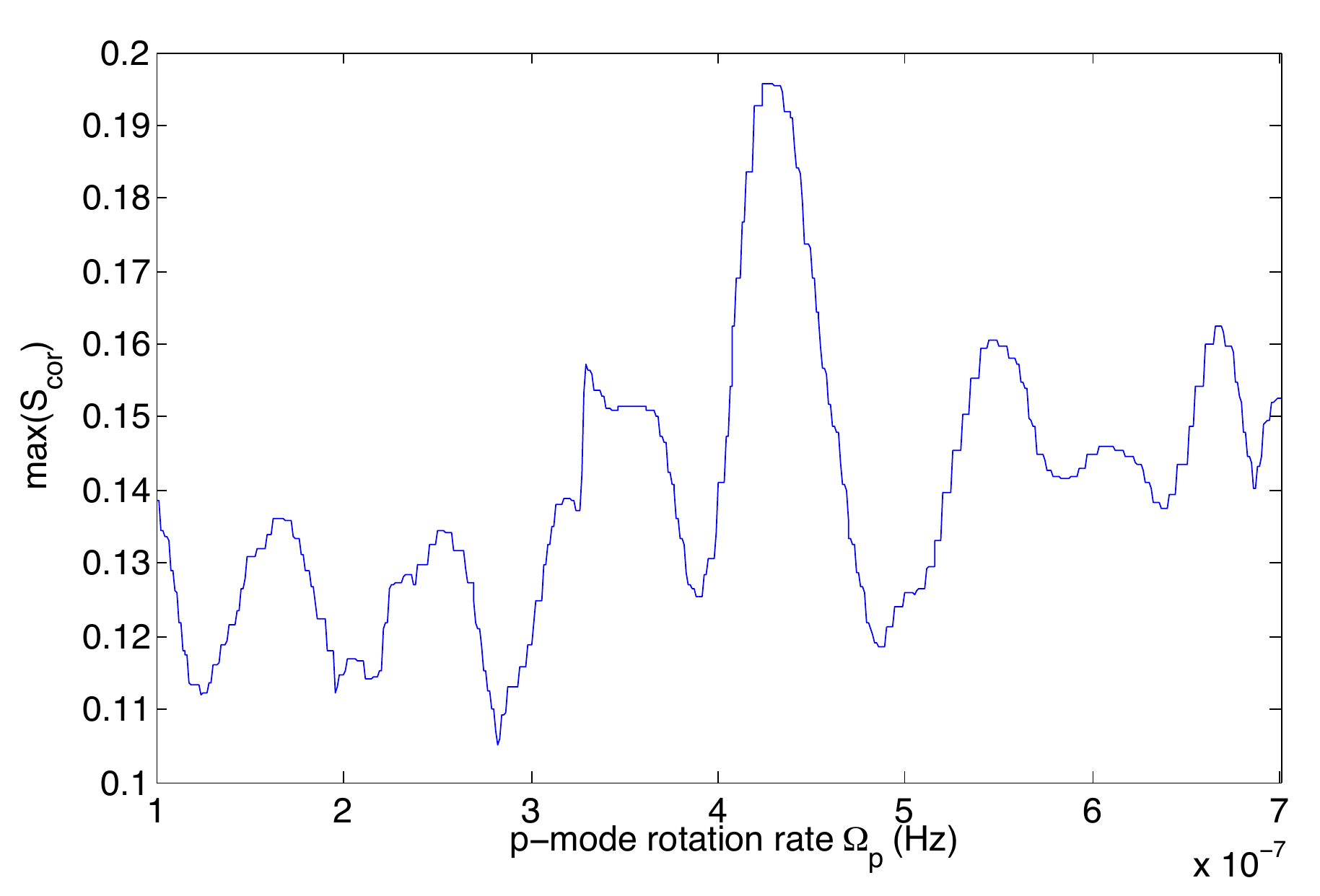}
\caption{Assumed unknown p-mode rotation rate tested against the optimum visibility of the three  g-mode splitting values that are assumed to exist, but are
unknown so far. The well-known 433~nHz p-mode rotation rate is confirmed by the g modes, only a little less accurately than when measured directly on the p modes themselves.}
\label{fig:maxScor}
\end{figure} 

\begin{figure}
\centering
\includegraphics[width=\columnwidth]{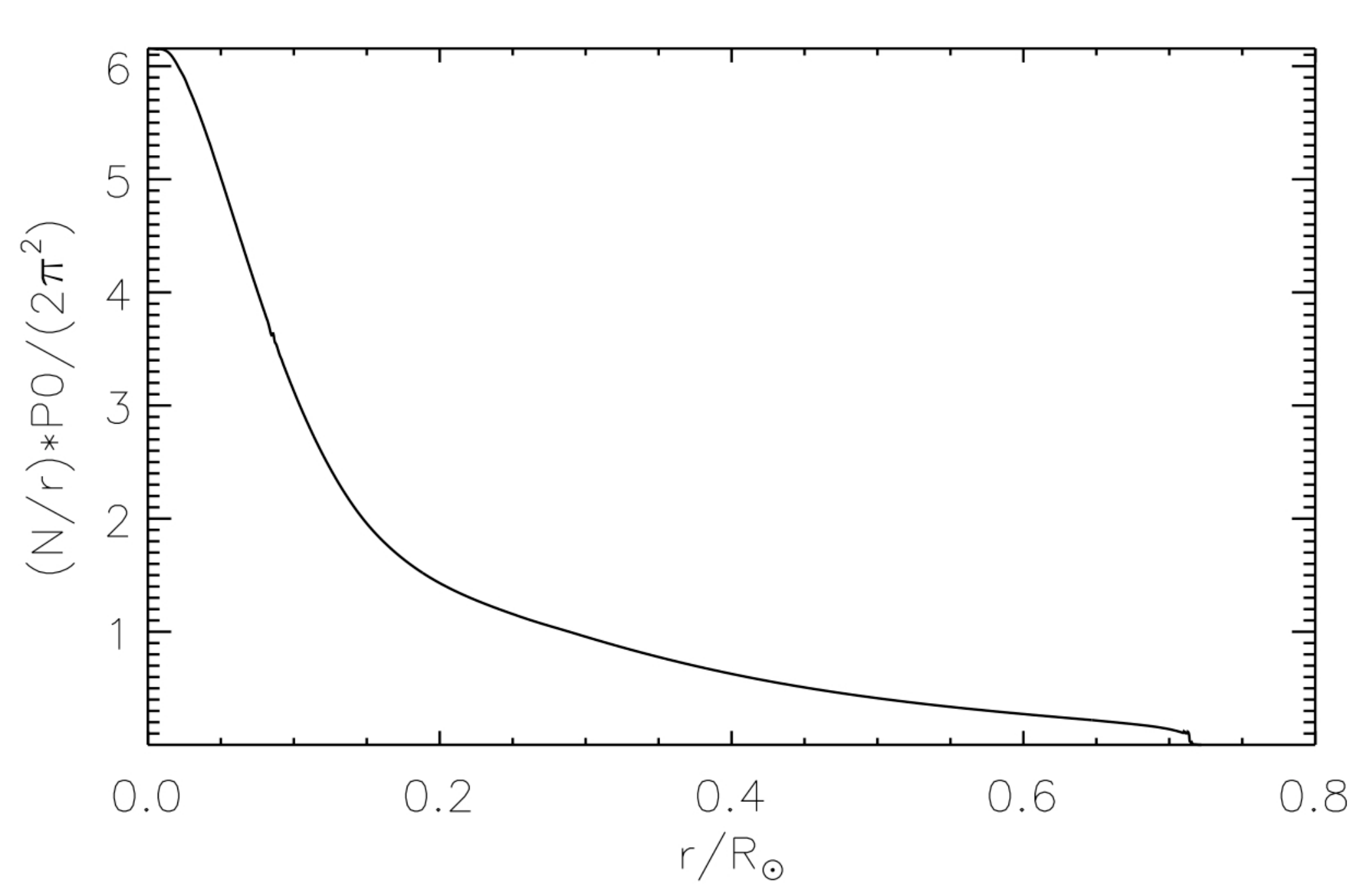}
\caption{Asymptotic g-mode rotational kernel $K(r)$ (Eq.~(\ref{eq:omegag})).}
\label{fig:K}
\end{figure}

\subsection{Equidistant period spacing of the asymptotic g-mode}

The broad frequency range across which this parameter has been measured from the dipole $l=1$ g modes implies an interesting accuracy on its value, as 76 equidistant modes have been exploited. The degree of freedom  for  targeting  the last mode from the first mode is extremely small. From the results displayed on Figs.~20 and 21, the equidistance can be estimated to be  $P_{1}=1443.1\pm 0.5$~s. 


Figure~21, produced by the sum of 76 equidistant (in period) symmetrical triplets, does not permit much doubt about this interpretation as a sequence of $l=1$ g modes. However, two different tests can be made to reinforce this assessment. \
The very high S/N ratio once again permits us to divide either the period range or the original time series itself into two independent parts that can then provide totally independent results.
Figures~31 and 32 illustrate these two possibilities. 

Figure~31 was obtained in the same way as Fig.~21, with the two independent time series of the first and second halves of the original 16.5 years. A factor of 0.7 was applied on the second half to display the two curves with a similar scale, as the background noise is about 30 percent higher in the second time series analysis. Despite an obvious reduction of the S/N ratio in comparison with Fig.~21, the first two periods of the series are still clearly visible and stand precisely at the same values of 30360 and 31804 s. We recall that these two analyses are made on completely independent data sets. 

\begin{figure}
\centering
\includegraphics[width=\columnwidth]{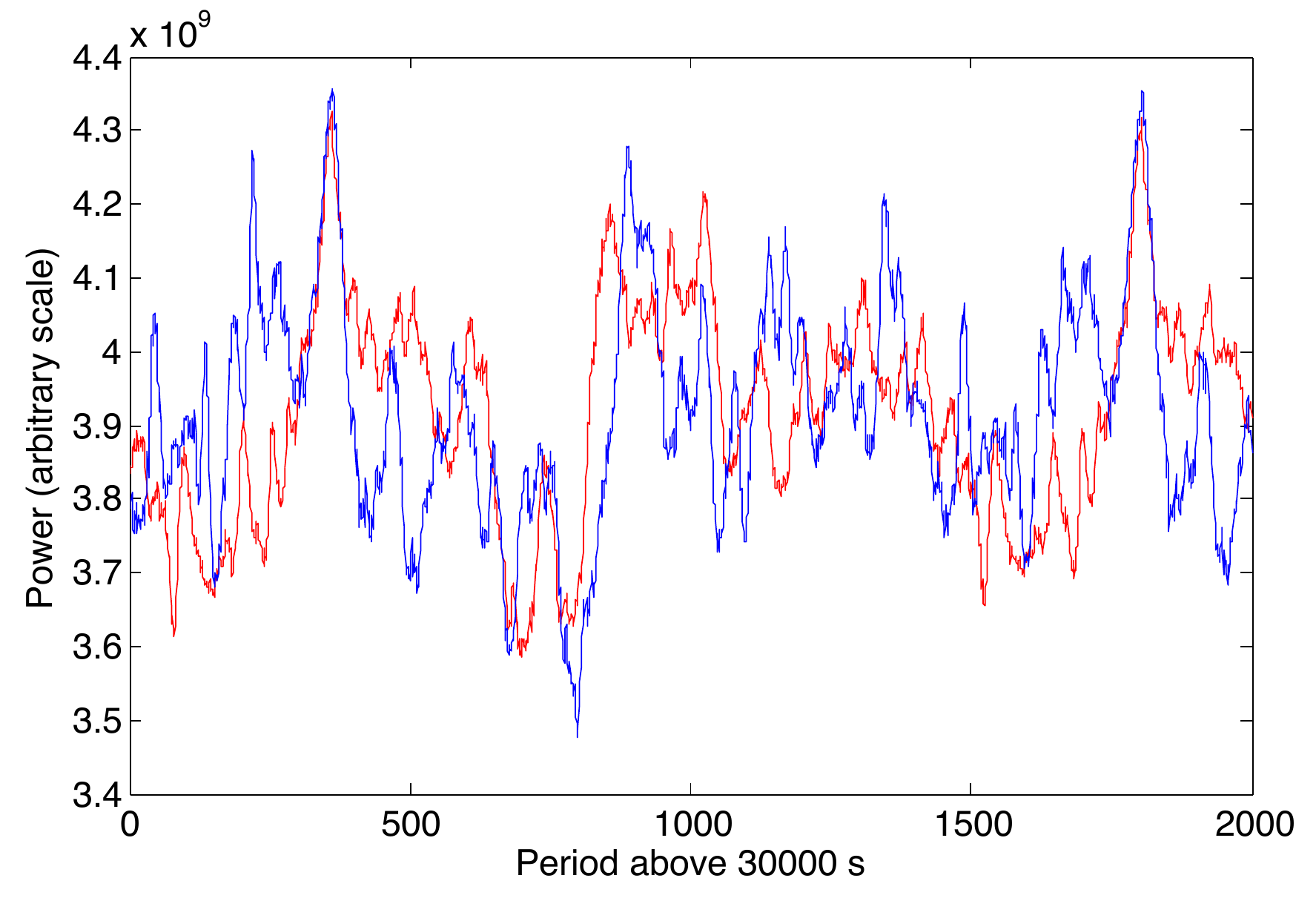}
\caption{Same as Fig.~21, computed on the same sequence of 76 layers of the power spectrum, equidistant in periods. This sum is computed separately here for the independent first half (red) and second half (blue) of the 16.5-year time series. The power scale of this and the next three figures is arbitrary, as they contain two curves that are rescaled to facilitate comparison.}
\label{fig:max_row_C1_8y_8y}
\end{figure} 

\begin{figure}
\centering
\includegraphics[width=\columnwidth]{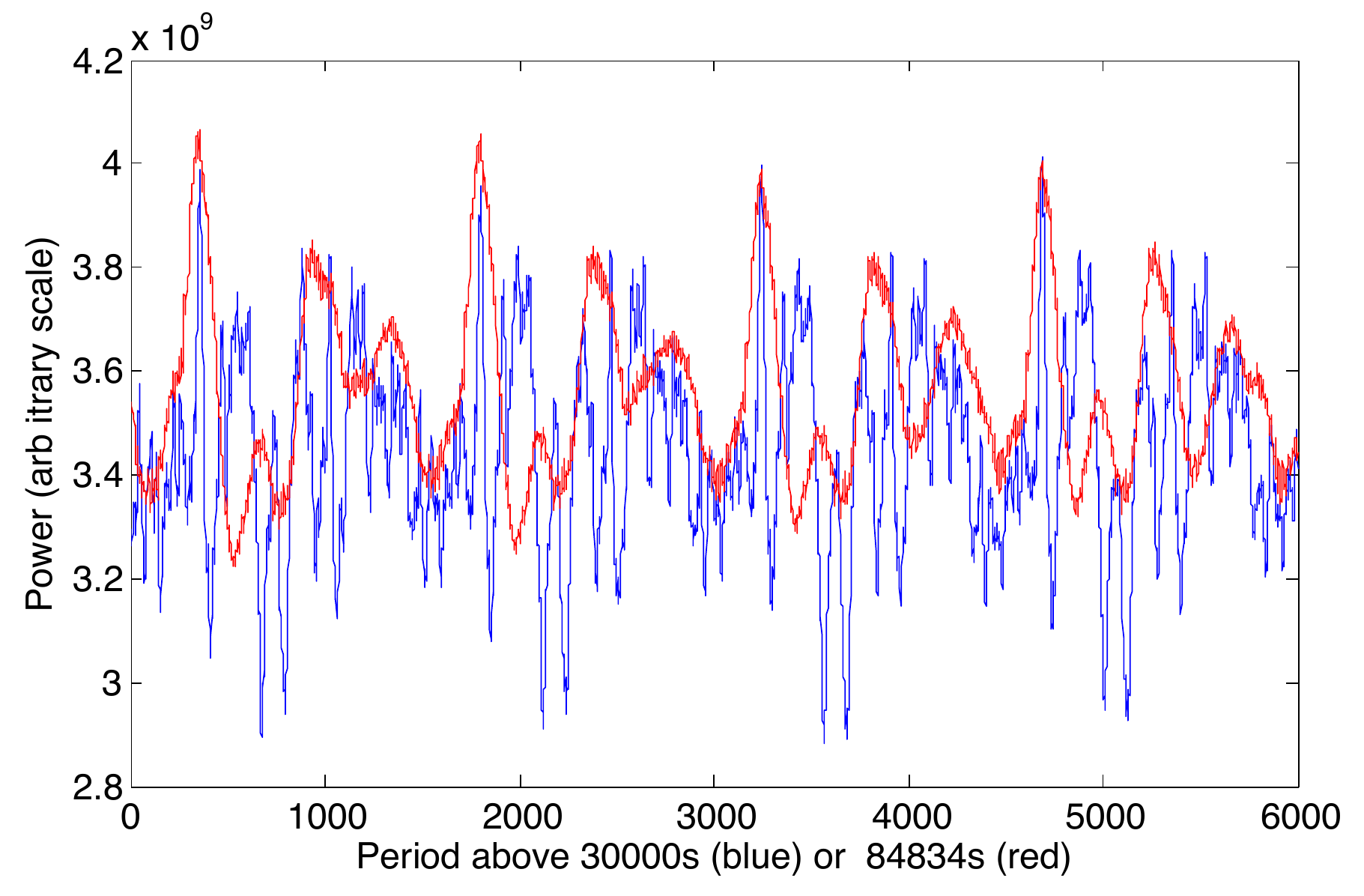}
\caption{Following Figs.~21 and 31, this illustration uses two independent sequences of 38 intervals that are equidistant in periods, the first half (shorter periods, radial order range from -20 to -57 for the first of the four peaks) plotted in blue, and the second half (longer periods, radial order range from -58 to -95 for the first peak, rescaled by a factor 0.8) shown
in red, both computed on the complete time series of 16.5 years. The first four periods of the sequences are shown. }
\label{fig:max_row_C1_20-57_58-95}
\end{figure}

In Fig.~32, the complete 16.5-year time series is used, but the sequence of 76 equidistant periods is separated into two. We used a first range that contains the first 38 periods with the same starting value of 30000 s, and another range that also contains 38 more periods, starting at 84834 s, which is 38 periods beyond the first range. This figure illustrates the first four periods of two series of 38 equidistant periods that are completely independent. Not only is the 1443~s periodicity clearly identified in these two independent parts, but the second periodicity is seen as an exact extension of the first. This gives confidence in the high accuracy of the period equidistance measurement, and it shows that  many g modes are indeed contributing in a very broad range of periods. We note that the second series (red) shows peaks that look significantly broader. This is due to the display in period, while the power spectrum itself is obtained in frequency. At lower frequencies, the same frequency bin contains a broader interval of periods.

The same two separations into independent time series and independent frequency ranges can be analyzed with the $l=2$ g modes. We have to expect noisier  results. The visibility is  indeed not as good as it is with $l=1$,  but as shown by Figs.~33 and 34, these separations clearly confirm that the $l=2$ modes are really present during the 16 years and also in the broad range of frequencies or periods. Consequently, this detection also validates the result
for the rotation of the solar core. 

\begin{figure}
\centering
\includegraphics[width=\columnwidth]{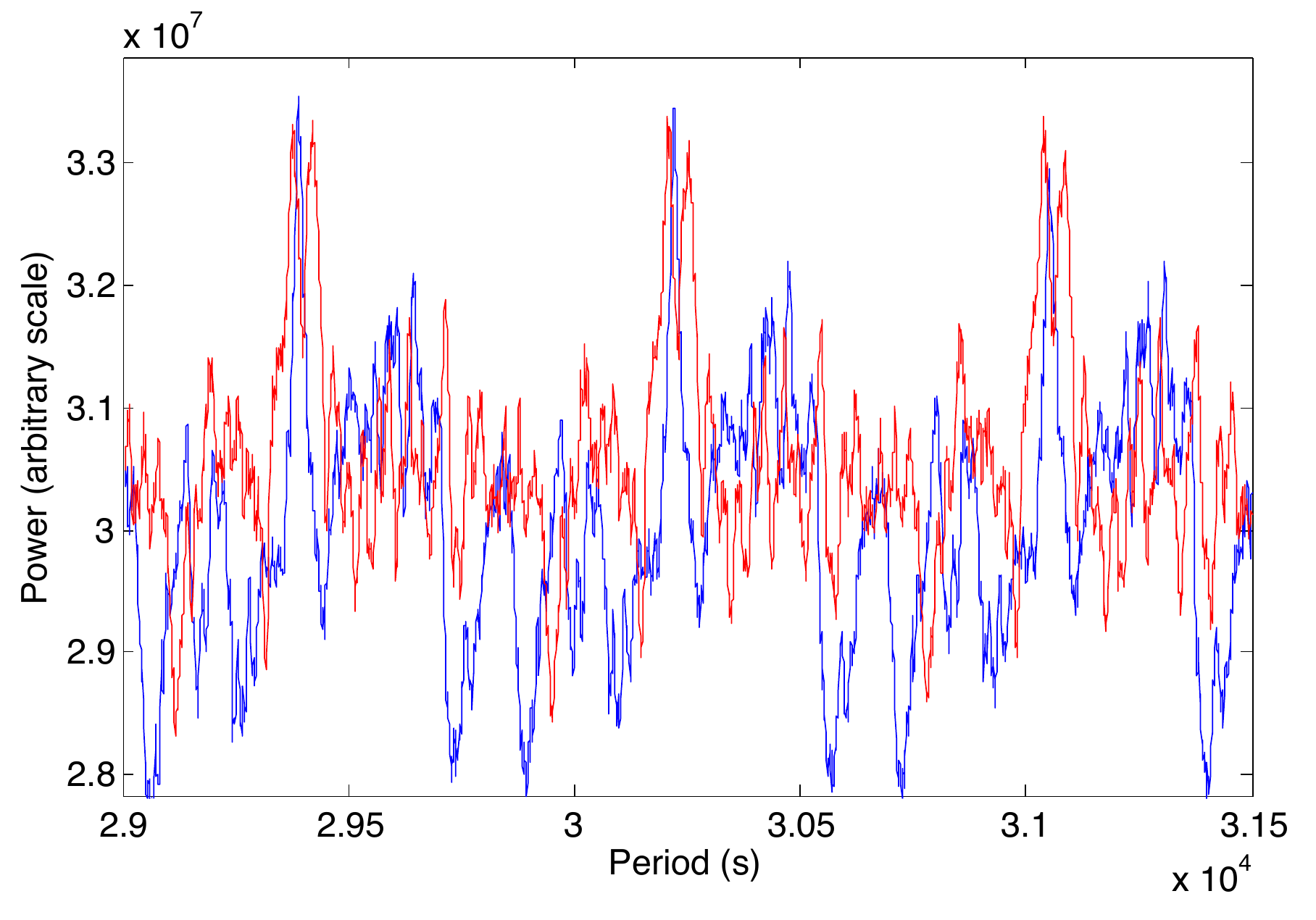}
\caption{This figure is to be compared to Fig.~27 for the $l=2$ g modes, computed for the same sequence of 112 layers of the power spectrum, equidistant in periods.  This sum is computed here separately for the independent first half (red) and second half (blue) of the 16.5-year time series. The first three periods of each sequence are shown.}
\label{fig:C2_cut_8y_8y}
\end{figure} 

\begin{figure}
\centering
\includegraphics[width=\columnwidth]{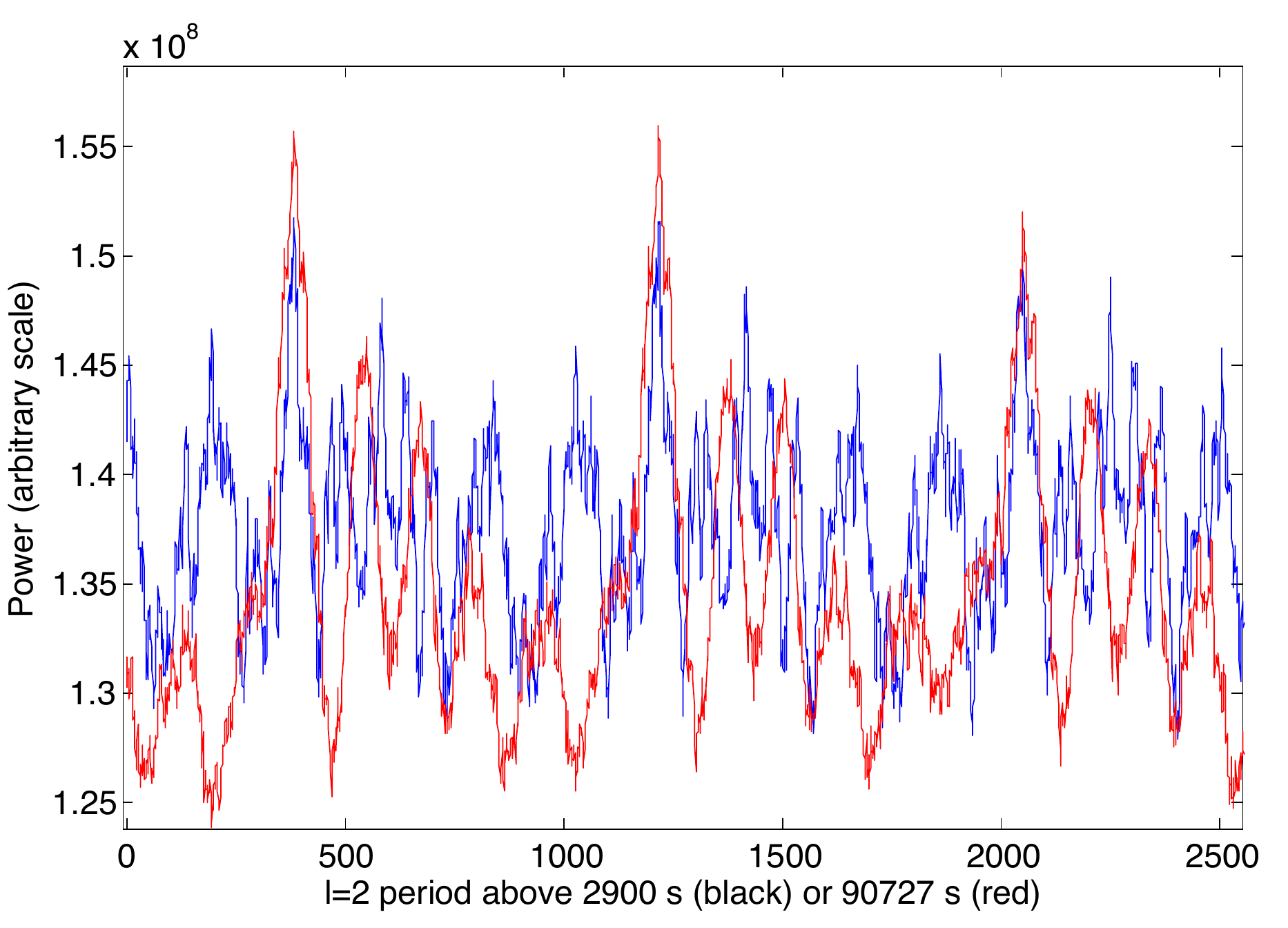}
\caption{As in Fig.~32 for $l=1$, this illustration for $l=2$ uses two independent sequences of 56 intervals equidistant in period, the first half (shorter periods, ranging from $n=-33$ to $-88$) is plotted in blue and the second half (longer periods, ranging from $n=-89$ to $-144$, slightly rescaled) is shown in red, both computed on the complete time series of 16.5 years. The first three periods of each sequence are shown.}
\label{fig:C2_cut_33-88_89-144}
\end{figure}

One last point has to be addressed in connection with this asymptotic g-mode equidistant period spacing: a possible departure from the asymptotic regime. All of the results shown so far assume a perfect equidistance in period, and their impressive sharpness supports this assumption. However, a small departure from this equidistance cannot be excluded and would slightly modify both the value of the real equidistance and the precise values of individual periods and frequencies. Provost \& Berthomieu~(1986) cautioned that at such accuracy of an equidistance measured within less than 1 second, we cannot pretend to have reached the asymptote as early as at $n = -20$. The difference between the periods and the linear asymptote should decrease following an hyperbolic trend of the radial order $n$, and the difference between consecutive periods should become constant within less than 1~s well beyond $n= -20$.
This was tested by taking a non-zero value of $\alpha$ in Eq.~(\ref{eq:model}), derived from Eq.~(\ref{eq:2ndasymptotic}), thereby 
introducing the hyperbolic decreasing term in the definition  of our 76-period model, 

\begin{equation}\label{eq:model}
P_{n,1} = P_{\mathrm{min},1} + P_1( |n|-20)+{\frac{\alpha}{|n|}}
.\end{equation} 

\begin{figure} 
\centering
\includegraphics[width=\columnwidth]{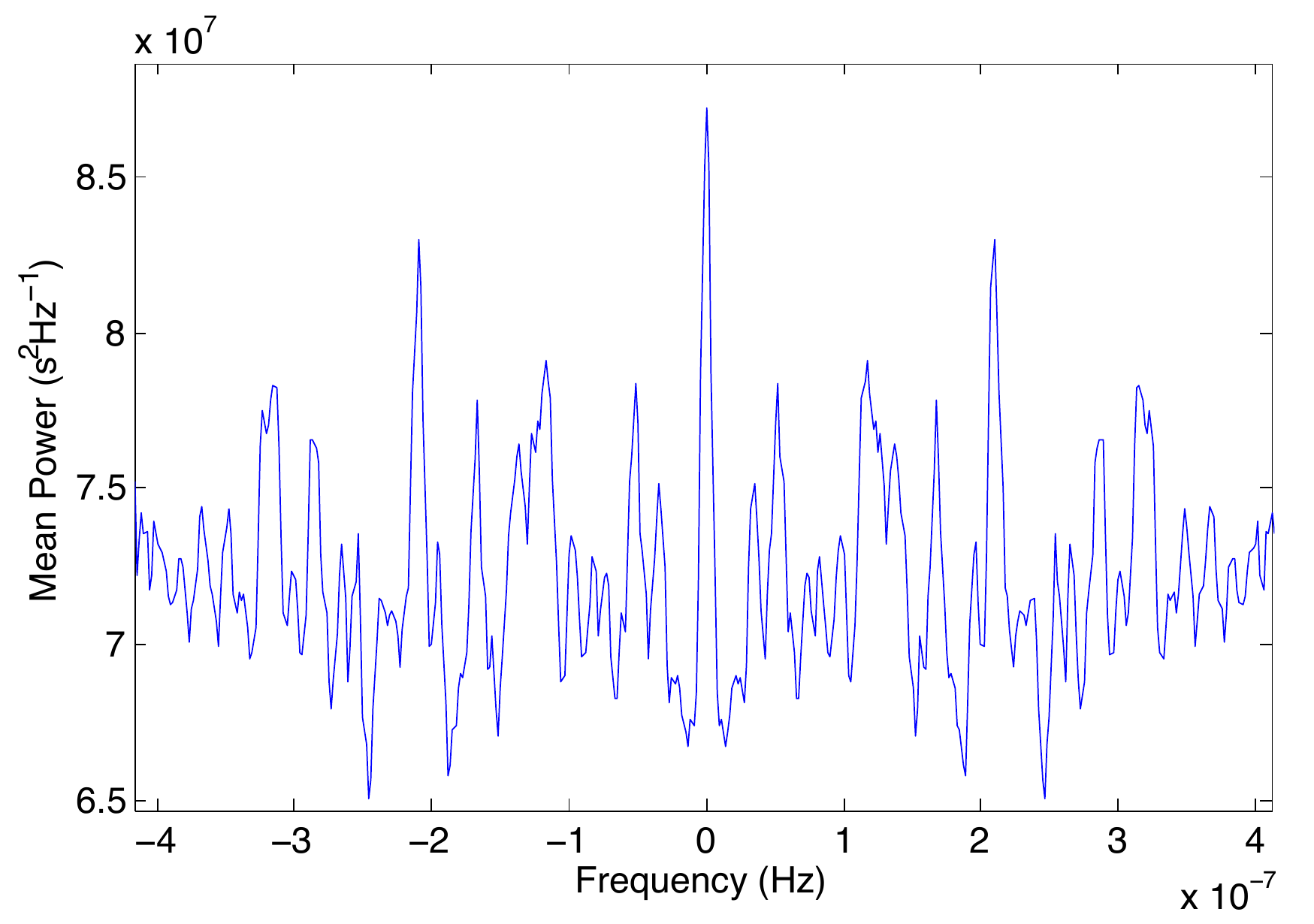}
\caption{Mean symmetrized profile of the 76 dipolar g modes,  following the model that includes a slight departure from the asymptotic equidistance. The comparison with Fig.~25 shows that taking this slight departure into account improves the visibility.}
\label{fig:M_sym}
\end{figure}

We then scanned a broad range of values of the parameter  $\alpha$ to look for a possible 
improvement of the S/N ratio of Fig.~21. The result is that, indeed, a small but significant improvement can be obtained with a positive value of  $\alpha$. The contrast of Fig.~21 increases by 6 percent with   $\alpha=1350$~s, and decreases irregularly on both sides of this value. This value of  $\alpha=1350$~s corresponds to an asymptotic equidistance of 1443.6~s and to a departure from this equidistance  of about 3~s at $n=-20$. It also implies a slight change in the period of the first identified g mode of radial order $n=-20$. A very precise adjustment can be made by optimizing the cumulated visibility of the mean triplet profile of these dipolar g modes. When we recompute the sum of the 76 spectral bands centered on the 76 frequencies defined by our model including this additional term  with  $\alpha=1350$~s, and then the average of this profile and its left-right flipped image, we have a sensitive tool for targeting the precise position of the central frequencies (if it is not precisely located, the central peak is not at zero in this plot and the average of the profile and its left-right image is clearly not optimal).
Incidentally, optimizing the visibility of the central peak simultaneously optimizes the combined visibility of the split side-lobes, which are the average of the two $m=\pm 1$ components. This means that these two split components are precisely symmetrical in frequency around the central peak. Figure~35 shows this mean symmetrized $l=1$ triplet profile, averaged over the 76 modes that are not quite equidistant in period. The splitting is clearly visible at its value of 209.5~nHz.
This fine analysis artificially equalizes the amplitudes of the $m=\pm 1$ tesseral components, but it has the additional benefit of improving the visibility of the triplet
by a factor $\sqrt2$ thanks to its precise left-right symmetry, and it brackets the value of the $n= -20$ period at $30385\pm 3$~s with $\alpha=1350$~s. However, this period and the equidistance are not independent, and both slightly depend on $\alpha$.

We conclude this discussion by setting the asymptotic equidistance at 1443.6~s with a conservative uncertainty of about 0.7~s. The period of the radial order -20 g mode, which is not yet on the asymptotic series, is then  $30385\pm 30$~s. Its distance from the asymptote is $1350/20=67.5$~s. 

In the case of $l=2$, the first accessible g modes correspond to radial orders  around 35. They are consequently closer to the asymptotic approximation, and no departure from this approximation can be detected by a similar analysis. The best choice for a similar $\alpha$ parameter is zero. The asymptotic equidistance remains  at 832.8~s with an uncertainty on the same order: 0.7~s. Assuming a perfect equidistance gives an uncertainty smaller than 0.7~s, but we have to take into account the small, but not measured, departure from the exact equidistance of the detected modes. For the same reason, even if we can measure the first period within a few seconds, this apparent accuracy assumes that everything is perfectly asymptotic. Taking the existence of a small but not detected slight departure from this linear approximation
into account, the true uncertainty on the first period of our model can again be estimated to 30 s. This is the number that we have mentioned above.

Taken together, the measured asymptotic equidistances of the
modes of degrees 1 and 2 provide a value of the asymptotic parameter $P_{0}$ (which is usually given in minutes rather than in seconds) $P_{0}=34$~mn~01~s$\pm 1$~s.

\subsection{About the g-mode amplitudes}

An interesting question arises concerning the amplitudes of the detected modes. This amplitude is not directly measured as a velocity of the waves' motions, it is indirectly detected by its effects on the sound waves' travel time integrated through the volume of the g-mode kernels. It is then expressed in seconds, to be taken as a fraction of this travel time. A careful calibration of the power spectrum shown in Fig.~7 and of all ensuing by-products such as the mean profile shown in Fig.~25 leads to a mean amplitude of 0.3~s for each individual component of each contributing dipolar g mode. This is a small fraction, on the order of 2$\times10^{-5}$, of the 4-hour travel time. It is less for the $l=2$ components, close to 0.1 s. These modulations occur only inside the g-mode kernels, however, and this means mostly inside the solar core itself where the sound waves travel much faster than in the radiative and convective envelopes. Its relative amplitude with respect to the travel time through the solar core is then rather on the order of 2-3$\times10^{-4}$, which is not so small, considering the many existing g modes that are revealed by this analysis. 

The meaning of this amplitude in terms of the frequency modulation rate on the p modes themselves remains to be investigated. It seems clear, however, that the search for g modes in the side lobes of the p modes could not give a similar result, either because the side lobes would have a tiny amplitude, much lower than the noise level between the modes, or because the nonlinear effects would spread the g-mode signature across many peaks, as noted by Lou~(2001), and would only contribute to the noise. 

When this is considered together with the rapid rotation of the deep solar layers implied by the values of g-mode splitting identified here, a new and very interesting challenge is raised for the study of solar core dynamics, as well as for physics itself. Previous studies dedicated to the individual search of some gravity modes have shown the potential of the Sun to place constraints on particle physics, including the search for dark matter in our Galactic environment (Turck-Chi\`eze \& Lopes 2012; Turck-Chi\`eze et al. 2012).

\section{Results and conclusions}

An original method has been applied to the search for solar g modes, based on an analysis of the round trip time of acoustic waves traveling through the solar diameter. This travel time is affected by temporal modulations, which are presumably caused by the presence of g modes that shake the structure of the core region. These modulations provide the signatures of many solar g modes with periods ranging between 0.4 and 2 days. This very low frequency range makes it possible to take statistical advantage of the asymptotic approximations, and thus to obtain extremely accurate results. A time series of 16.5 years of GOLF data was used for this analysis, and the results can also be obtained by separating this time series into two independent halves. This provides an extremely high level of confidence. The quantitative results include the asymptotic equidistance of g-mode periods, the periods themselves, and the rotation of the solar core.

 The mean rotation rate sensed by the asymptotic g modes, 
$\Omega_\mathrm{g}=1277\pm 10$~nHz, is a weighted average below the convection zone ($r\le r_\mathrm{cz}$). This leads to a mean value of the rotation rate  below $r_\mathrm{c}$, of 
$1644\pm 23$~nHz  (one-week period),  that is, a mean rotation of the solar core that is $3.8\pm 0.1$ times faster than the mean radiative zone rotation.

  This rapid rotation nevertheless remains difficult to explain by models describing a pure angular momentum evolution without adding new dynamical processes such as internal magnetic breaking, which could have appeared when the Sun was young (Turck-Chi\`eze et al. 2010).  Future more detailed analyses of the inversion will undoubtedly shed light on this intriguing aspect of solar dynamics. Clarifying the transition between a nearly solid rotation of the deep radiative envelope and a rapid rotation of the core just beneath this envelope will be an interesting challenge.

Besides the rotation rate of the solar core, the other quantitative parameters obtained by this analysis are the asymptotic period equidistances:

\begin{itemize}
\item for $l=l$, this is 
$P_{1}=1443.6~\mathrm{s} = 24~\mathrm{min}~03.6~\mathrm{s}\pm 0.7$~s, and
\item for $l=2$,  this is 
$P_{2}=832.8~\mathrm{s} = 13~\mathrm{min}~52.8~\mathrm{s}\pm 0.7$~s.
\end{itemize}
These two values taken together provide the parameter in Eq.~(\ref{eq:P0}):  
\begin{itemize}
\item 
$P_{0}=2041~\mathrm{s} = 34~\mathrm{min}~01~\mathrm{s}\pm 1$~s.
\end{itemize}
We have also proposed a tentative  identification of the shortest detected g-mode periods: 
\begin{itemize}  
\item for $l=1$, this is $P_{-20,1}=30385\pm30$~s, $\nu_{-20,1}=32.91\pm0.03\ \mu$Hz (very likely). 

The next 75  $l=1$ mode periods can be obtained by 

$P_{n,1} = P_{-20,1}+ (|n|-20)(P_1-67.5/|n|)$ for $-95\le n \le -21$.
\item for $l=2$, this is $P_{-33,2}=29380\pm30$~s, $\nu_{-33,2}=34.04\pm0.03\ \mu$Hz (likely).

The next 111 $l=2$ modes periods can be obtained by 

$P_{n,2}= P_{-33,2} + P_2(|n|-33)$  for $-144\le n \le -34$.
\end{itemize}

Our results in this very delicate and long quest have been obtained from two independent time series, as well as two independent frequency ranges. This lends them great credence. They open exciting new questions about the g-mode amplitudes, their excitation mechanism in such a broad and low-frequency range, and the solar rotation evolution. They provide strong new constraints for a dramatically improved description of the solar core.

\begin{acknowledgements}

The GOLF instrument on board SOHO is a cooperative effort of scientists, engineers, and technicians, to whom we are indebted.  For more details, see Gabriel et al 1995  or http://www.ias.u-psud.fr/golf/. GOLF data are available at MEDOC data and operations centre (CNES / CNRS / Univ. Paris-Sud), http://medoc.ias.u-psud.fr/. They are available as well on the GOLF web site. The data file used in this analysis has been calibrated  by R. Garc{\'{\i}}a. The high quality of the GOLF data is due equally to the outstanding performance of the SOHO platform and overall system. SOHO is a project of international collaboration between ESA and NASA. We would like to acknowledge the support received continuously during more than three decades from CNES. DS acknowledges the financial support from the CNES GOLF grant and the Observatoire de la C\^ote d'Azur for support during his stays. RKU acknowledges support from NASA for his participation in this
project and thanks John Bahcall for enthusiastic encouragement for
the g-mode search. We also wish to acknowledge R.M. Bonnet for his leadership in making the SOHO mission and GOLF a reality and P. Delache, who stimulated and motivated many of us in his broadly acknowledged role of g-mode ambassador. We wish to thank the referee for the initial suggestions that definitely helped to improve the general structure and to make the presentation more convincing.
\end{acknowledgements}
\bibliographystyle{aa}

\end{document}